\documentclass[journal,draftcls,onecolumn,11pt]{IEEEtran}

\usepackage{amsmath,graphicx,cite,amsfonts,dsfont,algorithm,algcompatible}
\usepackage{amssymb}
\algnewcommand\INPUT{\item[\textbf{Input:}]}%
\algnewcommand\OUTPUT{\item[\textbf{Output:}]}%
\algnewcommand\RETURN{\item[\textbf{Return:}]}%

\title{Hybrid Analog-Digital Transceiver Designs for Cognitive Large-Scale Antenna Array Systems}
\author{Christos G. Tsinos ~\IEEEmembership{Member,~IEEE,} Sina Maleki~\IEEEmembership{Member,~IEEE,}  \\ Symeon Chatzinotas~\IEEEmembership{Senior Member,~IEEE} and  Bj\"{o}rn Ottersten ~\IEEEmembership{Fellow,~IEEE} 
\thanks{The authors are with the Interdisciplinary
Centre for Security, Reliability and Trust (SnT), University of Luxembourg,
Luxembourg 2721, (Emails:\{christos.tsinos,sina.maleki,symeon.chatzinotas,bjorn.ottersten\}@uni.lu)}
\thanks{This work was supported by FNR, Luxembourg under the
CORE projects “SeMIGod” and “SATSENT” and by the European Commission through the “SANSA” project (ICT-645047).} 
}
\begin{document}
\markboth{SUBMITTED ON THE IEEE TRANSACTIONS ON SIGNAL PROCESSING}%
{SUBMITTED ON THE IEEE TRANSACTIONS ON SIGNAL PROCESSING}
\maketitle
\begin{abstract}

Milimeter wave (mmWave) band mobile communications can be a solution to the continuously increasing traffic demand in modern wireless systems. Even though mmWave bands are scarcely occupied, the design of a prospect transceiver should guarantee the efficient coexistence with the incumbent services in these bands. To that end, in this paper, two underlay cognitive transceiver designs are proposed that enable the mmWave spectrum access while controlling the interference to the incumbent users. MmWave systems usually require large antenna arrays to achieve satisfactory performance and thus, they cannot support fully digital transceiver designs due to high demands in hardware complexity and power consumption. Thus, in order to develop efficient solutions, the proposed approaches are based on a hybrid analog-digital pre-coding architecture. In such hybrid designs, the overall beamformer can be factorized in a low dimensional digital counterpart applied in the baseband and in an analog one applied in the RF domain. The first cognitive solution developed in this paper designs the cognitive hybrid pre-coder by maximizing the mutual information between its two ends subject to interference, power and hardware constraints related to the analog counterpart. The second solution aims at reduced complexity requirements and thus derives the hybrid pre-coder by minimizing the Frobenious norm of its difference to the optimal digital only one. A novel solution for the post-coder at the cognitive receiver part is further proposed here based on a hardware constrained Minimum Mean Square Error criterion. Simulations show that the performance of both the proposed hybrid approaches is very close to the one of the fully digital solution for typical wireless environments.
\end{abstract}
%
\section{Introduction}

Exponentially increasing demand for data rates as well as the spectrum congestion in the lower parts of the electromagnetic spectrum forces the wireless communications industry to explore systems adapted to frequencies within the so-called mmWave band \cite{BH,RSMZAWWSSG,OR1,OR2}. Mobile operators are investigating the possibility of using mmWave frequencies for future generation of mobile communications, while migrating the backhaul networks to beyond 30~GHz \cite{BH,RSMZAWWSSG}. Satellite industry has already established its presence in higher frequency bands by using the $18$~GHz band for feeder links as well as Fixed-Satellite-Service (FSS) user terminals in the downlink, and $28$~GHz band for the user terminal uplinks \cite{S}. Such a migration to tens of GHz bands provides the wireless industry with a much higher bandwidths, and thus enables them to deploy multiple Gbps services.  

However, the development of communications systems in the mmWave bands is a challenging task. MmWave signals suffer from severe propagation loss, penetration loss and rain fading compared to signals in lower frequencies \cite{MWLOSS}. Fortunately, the short wavelength of mmWave frequencies enables the denser packing of more antennas in the transceiver ends which now may employ large array structures for providing high beamforming gains. Furthermore, the large-scale transmit and receive arrays enable pre/post-coding techniques for multiple stream transmission resulting in significant spectral efficiency improvements. 

Even though conventional digital pre/post-coding techniques that were developed in the past years for lower frequency MIMO systems are independent of the carrier frequency, they cannot be applied in mmWave ones due to high demands in hardware complexity and power consumption. This is due to the one RF chain per antenna requirement for the implementation of a fully digital technique. Each RF chain includes a number of different electronic elements among which are Digital-to-Analog/Analog-to-Digital converters that have high requirements in hardware and power consumption. Thus, the implementation of a digital only mmWave transceiver is impractical and alternative approaches must be sought. 

A first solution to the problem was given via analog only beamforming approaches \cite{ANAL1,ANAL2,ANAL3,ANAL4,ANAL5}. The core of these techniques is a network of analog phase shifters \cite{PA1,PA2} that imposes constant modulus constraints on the beamformer and requires only one RF chain, which is highly desirable from a hardware complexity/power consumption point of view. Nevertheless, the major drawback of the analog only approaches is that they cannot support multi-stream communication. Therefore, in several cases they perform poorly compared to the fully digital approaches since they cannot exploit the inherent spatial multiplexing gain of the channel. An alternative approach replaces the phase shifters with even simpler components, analog switches \cite{AS1,AS2,AS3,AS4}. However, this results in loss in array gain which again leads to performance degradation, especially for highly correlated channels, like the ones experienced in the mmWave band.     

A more efficient solution is based on a two stage hybrid beamformer that consists of a low dimensional digital pre-coder applied in the BaseBand (BB) and an analog beamformer applied in the RF domain \cite{ARAPH}. The analog processing part is usually implemented via a network of phase shifters. This architecture can be applied to systems with limited the number of RF chains and achieve performance close to the one of a fully digital approach under some conditions. The concept of hybrid pre-coding was firstly introduced as the so-called ``antenna soft selection'' for point-to-point MIMO systems \cite{VPST,VPST2}. An interesting result in \cite{VPST} showed that for single stream transmission, a hybrid beamformer can realize the corresponding fully digital one provided that both the transmitter and receiver ends are occupied with at least two RF chains. Recent works \cite{MEND,SY} generalize this result to multi-stream communications by showing that the required number of chains at both transceiver ends must be two times the number of transmitted streams in order for a hybrid architecture to achieve the optimal digital only solution. 

In mmWave bands, a similar approach to the one of antenna soft selection was introduced by the seminal papers \cite{ARAPH,AALH}. In \cite{ARAPH}, it is shown that the spectral efficiency maximization problem can be closely approximated by minimizing the Frobenious norm of the difference of the fully digital solution to the product of the analog-digital counterparts of the hybrid pre-coder. 
The analog counterpart is selected from a pre-determined codebook and thus, the resulting problem is simplified by the recent advances in sparse reconstruction signal processing literature \cite{SP1,SP2}. The performance of this approach is satisfactory for large array systems that have in general strictly more RF chains than the transmitting streams and for low rank channels. In \cite{YSZL}, the Frobenious norm approximation problem of \cite{ARAPH} was solved without using a codebook for the analog counterpart solution. Their solution was based on an alternating minimization procedure which it is shown to achieve significantly better performance than the one of \cite{ARAPH} when the number of the available RF chains at both transceiver ends is equal to the number of the transmitting streams. In \cite{SY}, a codebook free approach is again proposed, though now the analog and digital parts of the beamformer are derived by solving two independent optimization problems after decoupling the original one. The decoupling of the analog-digital design is based on the assumption that the digital part is approximately orthonormal which in general holds for very large array systems. The performance of the latter approach is also superior compared to the one of \cite{ARAPH} for equal number of RF chains and transmitting streams.            

In the literature of mmWave communications cited above, the wireless system is assumed as a stand-alone MIMO communications system. Considering the fact that mmWave bands are also highly allocated to several services (e.g. point-to-point (P2P) or point-to-multipoint (P2MP) backhaul microwave links, satellite links, high resolution radar, radio astronomy, amateur radio e.t.c.) \cite{INC1,INC2}, such an assumption seems to be very optimistic. Given also the current 5G projections for traffic increase that imply an increase on backhaul demand \cite{RSMZAWWSSG}, new generation backhaul networks will have to coexist with older generations since traditional solutions like network planning cannot meet the demands. Therefore, it is necessary to take the constraints on the imposed interference on the incumbent services into account while designing the mmWave transceivers. While several digital only CR transceiver designs have been proposed in literature so far \cite{RZ,SPOOL,T,SHREE}, a mmWave system cannot afford to accommodate the number of RF chains that these techniques require for the reasons discussed above in this section. This in turn can be alleviated by designing hybrid analog-digital transceivers subject to interference temperature constraints as in the case of traditional underlay cognitive radios. This is the subject of research in this paper.


In detail, the contributions of the present paper are as follows. Two cognitive hybrid analog/digital transceiver designs for point-to-point MIMO systems are presented with the view to maximize the spectral efficiency subject to interference to the Primary User, power and hardware related (limited number of RF chains) constraints. Both of the proposed approaches are based on the underlay cognitive radio paradigm \cite{GOLD} where the Secondary User (SU) may access a spectrum area licensed to a Primary User (PU), simultaneously with the latter and provided that the interference power on the PU transmissions is below a predefined threshold. The proposed approaches derive the pre-coding and post-coding (combining) matrices by decoupling the transmitter receiver optimization problem. Thus, at first the pre-coder is derived such that the system's mutual information is maximized subject to the constraints mentioned previously in this paragraph and then for that pre-coder, the optimal linear post-coding matrix is derived in a Minimum Mean Square Error (MMSE) sense. The proposed transceiver designs are not based on a codebook and involve the solution of difficult non-convex optimization problems. The developed algorithms for their solution are based on the so-called Alternating Direction of Multipliers Method (ADMM) \cite{AGL} which is known for its good behaviour in several non-convex optimization problems \cite{ADMM}, similar to the ones that are dealt in the present paper. 

The first approach derives the analog and digital pre-coding matrices by solving directly the  mutual information maximization problem.
As it was discussed before, existing solutions are based on  approximations of the original problem or on a decoupled version with respect the analog and digital counterparts. As a consequence, the solution derived in the present paper achieves satisfactory performance under any channel model and not only for low rank/highly correlated channels, similar to the ones that are based on the parametric mmWave models. Thus, it may also be successfully applied in large array systems that function in lower frequencies and under typical fading channel models in order to reduce the hardware complexity/power consumption. Moreover, it can be further applied to systems that have relative small number of antennas since it does not require orthonormal properties to be satisfied by the digital pre-coder part, as it is the case for the designs that are based on the decoupled analog-digital problem. 

While the first hybrid solution presents excellent performance under any conditions, its complexity is rather high due to the $\log_2\det(\cdot)$ function that appears in the cost function of the mutual information maximization problem. To overcome this, we follow the approach of \cite{ARAPH} and approximate the solution to the mutual information maximization problem by the one that minimizes the Frobenious norm of the difference of the hybrid beamformer to the optimal digital only one. A novel algorithm that requires significantly reduced complexity is derived to provide the solution to the last problem while it achieves close performance to the one of the first hybrid solution (mutual information maximization problem) for typical mmWave environments.

Furthermore, a novel solution for the cognitive receiver side post-coder is also   developed. An efficient code-book free hybrid analog/digital post-coder is derived based on the MMSE criterion and subject to the hardware constraints related to the phase shifters of the analog processing network. On the contrary, existing approaches are either codebook based \cite{ARAPH} or are considering the codebook free design of only the hybrid pre-coding matrix at the transmitter side \cite{YSZL,SY}.  

The convergence properties of the proposed algorithms are theoretically studied and their performance is evaluated via simulations for typical channel models.  
         
The remainder of the paper is organized as follows. In  Section II, the system description is given. in Section III, the description of the optimal digital only solutions is given. Section IV describes the hybrid transceiver design based on the mutual information maximization criterion. In Section V, the alternative hybrid pre-coder design which is based on the Frobenious norm approximation criterion, is presented. Section VI presents some numerical results and Section VII concludes the work.

\section{System Model}
\label{SEC:SYS_DEC}

Let us assume that a Secondary User (SU) $R_s \times T_s$ MIMO system access a spectrum area allocated to a $R_p \times T_p$ Primary User (PU) by employing a typical cognitive underlay approach (Fig. 1) \cite{GOLD}. 
\begin{figure}[h]
    \centering
    \includegraphics[width=0.5\textwidth]{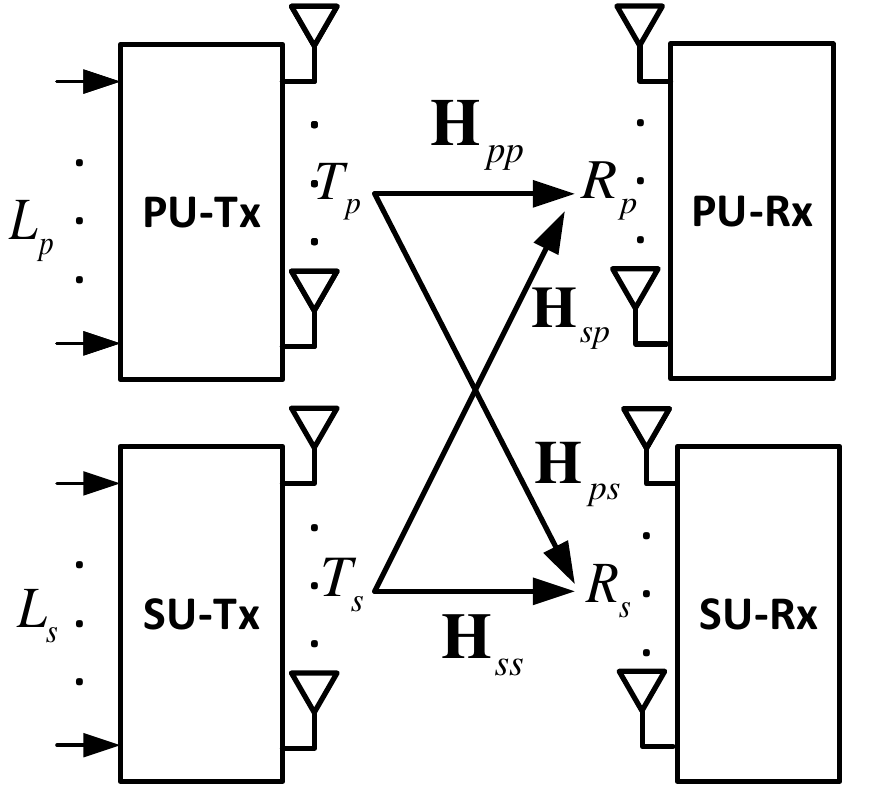}
    \caption{PU-SU transceiver pairs. The SU is equipped with $N_{st} << T_s$ and $N_{sr} << R_s$ RF chains at the transmitter's and receiver's side, respectively.}
    \label{FIG:SYS_MOD}
\end{figure}
The SU system (Fig. 2) is assumed to be equipped with $N_{st} << T_s$ and $N_{sr} <<R_s $ RF chains at the 
transmitter and the receiver, respectively and transmits a number of  $L_s \leq \min\{N_{st},N_{sr}\}$ streams. Moreover, the SU system applies a $T_s \times L_s$ hybrid 
pre-coding matrix at the transmitter given by $\mathbf{F} = \mathbf{F}_{RF}\mathbf{F}_{BB}$ where $\mathbf{F}_{RF}$ is the $T_s \times N_{st} $ analog RF 
precoder implemented via a phase shift network and $\mathbf{F}_{BB}$ is the $N_{st} \times L_s$ digital BB one. In a similar manner, the $R_s \times L_s $ 
post-coding matrix is given by $\mathbf{W} = \mathbf{W}_{RF}\mathbf{W}_{BB}$ where $\mathbf{W}_{RF}$ is the $R_s \times N_{sr} $ analog post-coding matrix and $\mathbf{W}_{BB}$ is the $N_{sr} \times L_s$ digital BB one, respectively.

The received signal $\mathbf{y}_s$ at the SU receiver before and after post-coding, assuming a narrow-band block fading propagation channel is given by
\begin{align}
\mathbf{y}_s &= \mathbf{H}_{ss}\mathbf{F}\mathbf{x}_s+ \tilde{\mathbf{H}}_{sp}\mathbf{x}_p+\mathbf{n}_s \\
\mathbf{y}_s' &= \mathbf{W}^H\mathbf{y}_s  
\label{EQ:SU_INOUT} 
\end{align}
where $\mathbf{H}_{ss}$ is the $R_s \times T_s$ channel matrix between the two SU ends, $\mathbf{x}_s$ is the $L_s \times 1$ vector of the SU transmitted symbols such that $\mathbb{E}\{\mathbf{x}_s\mathbf{x}_s^H\} = \sigma_s^2\mathbf{I}_{T_s}$, $\mathbb{E}\{\cdot\}$ is the expectation operator, $\sigma_s^2$ is the variance of the symbols transmitted by the SU transmitter, $\mathbf{I}_{T_s}$ is the $T_s \times T_s$ identity matrix,  $\tilde{\mathbf{H}}_{sp} = \mathbf{H}_{sp}\mathbf{F}_p$, $\mathbf{H}_{sp}$ is the $R_s \times T_p$ channel matrix between the PU transmitter and the SU receiver and $\mathbf{F}_p$ is the $T_p \times L_p$ pre-coding matrix applied at the PU transmitter, $L_p\times 1$ is the number of the transmitted PU streams, $\mathbf{x}_p$ is the $L_p \times 1$ vector of the PU transmitted symbols such that $\mathbb{E}\{\mathbf{x}_p\mathbf{x}_p^H\} = \sigma_p^2\mathbf{I}_{T_p}$, $\sigma_p^2$ is the variance of the symbols transmitted by the PU transmitter and 
\begin{figure}[htbp]
    \centering
    \includegraphics[width=0.9\textwidth]{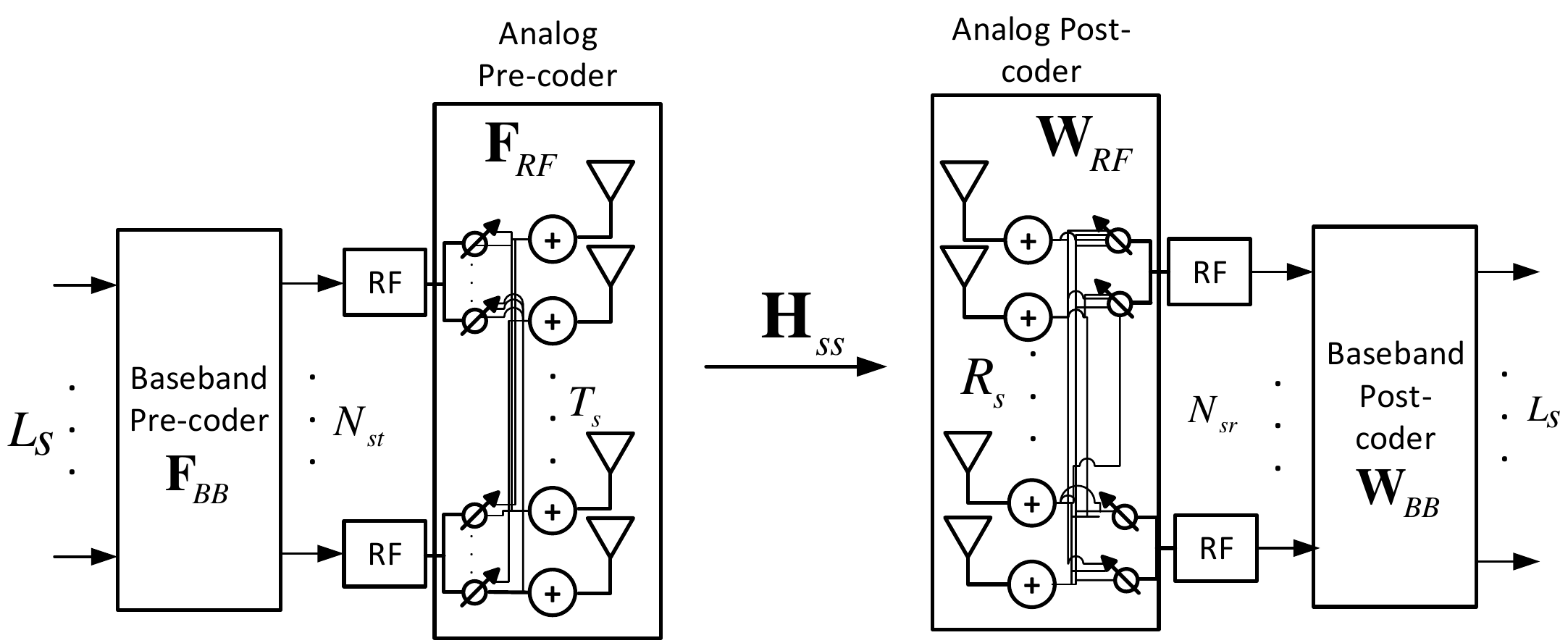}
    \caption{SU Hybrid Analog/Digital Transceiver}
    \label{FIG:I}
\end{figure}
$\mathbf{n}_s$ is i.i.d. complex Gaussian noise modeled as $\mathcal{CN}(0,\sigma_n^2\mathbf{I}_{R_s})$.

Note that the SU does not require explicit knowledge of matrix $\mathbf{F}_p$ since it requires knowledge of only the overall matrix $\tilde{\mathbf{H}}_{sp}$. Moreover, the PU system may also apply a hybrid pre-coding technique in order to optimize its transmission or not (i.e conventional MIMO system), though  as the proposed technique is independent of the latter, we chose an abstracted description to simplify the analysis.    

%

In a similar manner, the received signal at the PU receiver is given by
\begin{equation}
\mathbf{y}_p= \mathbf{H}_{pp}\mathbf{F}_p\mathbf{x}_p+ \mathbf{H}_{ps}\mathbf{F}\mathbf{x}_s+\mathbf{n}_p,
\label{EQ:PU_INOUT} 
\end{equation}                    
where $\mathbf{H}_{pp}$ is the $R_p \times T_p$ channel matrix between the two PU ends, $\mathbf{H}_{ps}$ is the $R_p \times T_s$ channel matrix between the SU transmitter and the PU receiver and $\mathbf{n}_p$ is i.i.d. complex Gaussian noise modeled as $\mathcal{CN}(0,\sigma_n^2\mathbf{I}_{R_p})$. 

The mmWave channel presents limited spatial diversity due to the high free space pathloss and high antenna correlation due to the dense packing of large number of antennas. A geometrical channel model captures better these characteristics of the mmWave band. Thus, by assuming that each of the involved systems employ an Uniform Linear Array (ULA) of antennas and $N_p$ propagation paths are between a transmitter $a=\{s,p\}$ and a receiver  $b=\{s,p\}$, the discrete narrow-band channel is given by \cite{SY}\cite{LXD}
\begin{align}
\mathbf{H}_{ba} = \sqrt{\frac{T_aR_b}{N_{p}}} \sum_{l=1}^{N_{p}}\alpha_{l}\mathbf{a}_{br}(\phi_{l}^{br})\mathbf{a}_{at}(\phi_{l}^{at})^H,
\end{align} 
where $\alpha_{l} \sim \mathcal{CN}(0,1)$ is the complex gain of the $l$th path, $\phi_{l}^{br}$  and $\phi_{l}^{at}$ are the azimuth angles of arrival and departure, respectively, that are uniformly distributed in $[0, 2\pi)$. Vectors $\mathbf{a}_{br}(\phi_{l}^{br})$ and $\mathbf{a}_{at}(\phi_{l}^{at})^H$ are the normalized receive and transmit response vectors at the azimuth angle of arrival and departure $\phi_{l}^{br}$ and $\phi_{il}^{at}$, respectively. Furthermore, the response of an ULA of $T$ antenna elements is given by
%
%
\begin{equation}
\mathbf{a}(\phi) = \frac{1}{\sqrt{T}}\left[1, e^{jkd\sin(\phi)}, \dots, e^{j(T-1)kd\sin(\phi)}\right],
\end{equation}
where $k= \frac{2\pi}{\lambda}$, $\lambda$ is the carrier wavelength and $d$ is the inter-element spacing.

Let us now assume that the SU has perfect knowledge of $\mathbf{H}_{ss}$, $\tilde{\mathbf{H}}_{sp}$ and $\mathbf{H}_{ps}$ channels. Note that this is a typical requirement in CR underlay approaches \cite{GOLD}. In general, since both the SU and PU systems are operating in the same frequency and under the channel reciprocity assumption, it is straightforward to estimate the required information related to the cross SU-PU channels via the training symbols of the PU system. Since this a matter well-studied within the CR literature \cite{GOLD,RZ,SPOOL,T} we avoid any further discussion. The aim is to derive the pre/post-coding matrices in order to maximize the spectral efficiency of the SU system subject to constraints on the total transmission power ($P_{max}$) and on the interference to the PU transmissions ($I_{max}$). To that end, the SU system must solve the following optimization problem
\begin{align}
\hspace{-25pt}(\mathcal{P}_1): \hspace{25pt} &\max_{\mathbf{F},\mathbf{W}} R(\mathbf{F},\mathbf{W}) \nonumber \\
&s.t. \hspace{10pt} tr(\mathbf{F}\mathbf{F}^H) \nonumber \leq P_{max} \\
& \hspace{25pt} tr(\mathbf{H}_{ps}\mathbf{F}\mathbf{F}^H\mathbf{H}_{ps}^H)\leq I_{max}
\label{EQ:SU_OPT1}
\end{align} 
where the spectral efficiency $R(\mathbf{F},\mathbf{W})$ is defined as 
\begin{equation}
R(\mathbf{F},\mathbf{W}) = \log_2\det\left(\mathbf{I}_{L_s}+\mathbf{R}_n^{-1}\mathbf{W}^H\mathbf{H}_{ss}\mathbf{F}\mathbf{F}^H\mathbf{H}_{ss}^H\mathbf{W}\right),
\label{EQ:SU_SE} 
\end{equation} 
with $\mathbf{R}_n= \mathbf{W}^H(\tilde{\mathbf{H}}_{sp}\tilde{\mathbf{H}}_{sp}^H+\sigma_n^2\mathbf{I}_{L_s})\mathbf{W}$ denoting the interference plus noise covariance matrix at the SU receiver and $\det(\cdot)$ is the determinant of a matrix. It is instructive to note that by imposing the interference power constraint on \eqref{EQ:SU_OPT1}, the capacity loss of the PU due to the SU transmissions is well regulated as it is proved in \cite{RZ}. Thus, by setting $I_{max}$ to a sufficiently small value with respect to the noise power $\sigma_n^2$, the aforementioned capacity loss can be made arbitrary small.

For the sake of simplicity, we consider in this paper the case where interference constraints are posed from only one PU. Extensions to the multiple PU case are straightforward by adding the corresponding interference related constraints on $(\mathcal{P}_1)$.  

Prior to presenting the proposed hybrid pre/post-coding solutions, the equivalent digital only solution is discussed which will serve as a benchmark to the performance of the aforementioned hybrid approaches.

\section{Digital Only Solution}   

In the digital only design, the SU pre/post-coding matrices can be found by solving directly $(\mathcal{P}_1)$ under the assumption that both the transmitter and the receiver ends have equal number of RF chains and antennas. This problem is in general intractable \cite{POLO} and the common approach is to temporally decouple the designs at the transmitter and the receiver. To that end, the optimal pre-coding matrix $\mathbf{F}_D$ is designed such that the mutual information achieved by Gaussian signaling over the wireless channel is maximized. Therefore, the following optimization problem should be solved
\begin{align}
\hspace{-10pt}(\mathcal{P}_2): \hspace{10pt} &\max_{\tilde{\mathbf{F}}} \log_2\det(\mathbf{I}_{L_s}+\mathbf{Q}^{-1/2}\mathbf{H}_{ss}\tilde{\mathbf{F}}\mathbf{H}_{ss}^H\mathbf{Q}^{-1/2}) \nonumber \\
&s.t. \hspace{10pt} tr(\tilde{\mathbf{F}}) \leq P_{max} \nonumber \\
& \hspace{25pt} tr(\mathbf{H}_{ps}\tilde{\mathbf{F}}\mathbf{H}_{ps}^H)\leq I_{max} \nonumber \\
& \hspace{25pt} \tilde{\mathbf{F}} \succeq \mathbf{0}
\label{EQ:SU_OPT_DIG_F}
\end{align} 
where $\mathbf{A} \succeq \mathbf{0}$ denotes a positive semi-definite matrix, $\mathbf{Q} = \tilde{\mathbf{H}}_{sp}\tilde{\mathbf{H}}_{sp}^H + \sigma_n^2\mathbf{I}_{R_s}$ is the covariance matrix of the interference plus noise signal and $\tilde{\mathbf{F}} = \mathbf{F}_D\mathbf{F}_D^H$. In fact, $\mathbf{Q}^{-1/2}$ is a whitening post-processing filter that maximizes the mutual information between the input and the output of the SU system \cite{WHITE}. Note that $(\mathcal{P}_2)$ is convex and thus, it can be solved efficiently by standard convex optimization techniques (i.e. interior point methods \cite{BOYD}). Then, by applying the Eigenvalue Value Decomposition on the optimal matrix $\tilde{\mathbf{F}}^*=\mathbf{U}_F\mathbf{\Sigma}_F\mathbf{U}_F^H$, we may find $\mathbf{F}_D=\mathbf{U}_F\sqrt{\mathbf{\Sigma}_F}$.  

Problem $(\mathcal{P}_2)$ aims at maximizing the SU's channel capacity assuming that the SU receiver has full CSI and can perform optimal nearest-neighbor decoding based on the $R_s$ dimensional received signal \cite{GOLDII}. This decoding approach requires high computational complexity even for typical MIMO systems that have very few antennas compared to a mmWave one. A common approach that avoids the high complexity of the previous solution is to use a linear receiver design. To that end, in this paper a typical Minimum Mean Square Error (MMSE) receiver is considered for the receive side of the SU system. Thus,
matrix $\mathbf{W}$ is derived as the solution to the optimization problem
\begin{equation}
\min_{\mathbf{W}} \mathbb{E}\left\{\|\mathbf{x}_s - \mathbf{W}^H\mathbf{y}_s\|_F^2\right\},
\label{EQ:MMSE}
\end{equation}
%
%
where $\|\cdot\|_F$ denotes the Frobenius norm. The solution to the previous problem is given by \cite{KAH}
\begin{align}
\mathbf{W}_{MMSE} =& \mathbb{E}\{\mathbf{y}_s\mathbf{y}_s^H\}^{-1}\mathbb{E}\{\mathbf{x}_s\mathbf{y}_s^H\}^H \nonumber \\
=& \left(\mathbf{H}_{ss}\mathbf{F}_D\mathbf{F}_D^H\mathbf{H}_{ss}^H+\tilde{\mathbf{H}}_{sp}\tilde{\mathbf{H}}_{sp}^H+\sigma_n^2\mathbf{I}_{R_s}\right)^{-1}\mathbf{H}_{ss}\mathbf{F}_D
\label{EQ:SU_OPT_DIG_W1}
\end{align}
where we used  \eqref{EQ:SU_INOUT}-\eqref{EQ:PU_INOUT} and the assumption that $\mathbb{E}\{\mathbf{x}_s\mathbf{x}_p^H\}=\mathbf{0}_{L_s \times L_p}$, where $\mathbf{0}_{L_s \times L_p}$ is a $L_s \times L_p$ matrix of zero entries. 

\section{Hybrid Analog/Digital Transceiver Design}
In this section first, the hybrid analog-digital transmitter is developed by solving the mutual information maximization problem $(\mathcal{P}_1)$ after casting it to the hybrid case. Then, the hybrid MMSE based solution for the post-coding matrix at the receiver is presented . 

\subsection{Transmitter Hybrid Design}
\label{SEC:HT} 

In the case of the hybrid approaches, the solution to the SU's spectral efficiency maximization problem $(\mathcal{P}_1)$ is even harder to be derived since the constraints related to the phase shifters network force the analog pre-/post-coding matrices to lie in the non-convex space of constant modulus complex matrices. In order to achieve a simple solution, the transceiver design should be again temporally decoupled. 

Let us first present the pre-coder design by considering the mutual information maximization problem $(\mathcal{P}_2)$ which is now cast to the hybrid pre-coding case, that is   
\begin{align}
\hspace{-2pt}(\mathcal{P}_3): \hspace{2pt}  &\max_{\mathbf{F}_{RF},\mathbf{F}_{BB}} \log_2\det(\mathbf{I}_{L_s}+\tilde{\mathbf{H}}_{ss}\mathbf{F}_{RF}\mathbf{F}_{BB}\mathbf{F}_{BB}^H\mathbf{F}_{RF}^H\tilde{\mathbf{H}}_{ss}^H) \nonumber \\
&s.t. \hspace{10pt} \|\mathbf{F}_{RF}\mathbf{F}_{BB}\|_F^2 \leq P_{max} \nonumber \\
& \hspace{25pt} \|\mathbf{H}_{ps}\mathbf{F}_{RF}\mathbf{F}_{BB}\|_F^2 \leq I_{max} \nonumber \\
& \hspace{25pt} \mathbf{F}_{RF} \in \mathcal{F}
\label{EQ:SU_OPT_ANAL_F}
\end{align} 
where $\tilde{\mathbf{H}}_{ss} = \sigma_s\mathbf{Q}^{-1/2}\mathbf{H}_{ss}$  and 
$\mathcal{F} = \left\{\mathbf{A} \in \mathbb{C}^{M \times N} \ | \ |\mathbf{A}(m,n)| = 1, 1 \leq m \leq M, \ 1\leq n \leq N \right\}$ for an arbitrary matrix $\mathbf{A}$ with complex entries $\mathbf{A}(m,n)$. Problem $(\mathcal{P}_3)$ is non-convex due to the bi-convex cost function and the non-convex constraint set $\mathcal{F}$. In the following, an efficient solution is derived by employing the so-called ADMM \cite{ADMM}. This method is an extension of the augmented Lagrangian multiplier method \cite{AGL} which first performs an alternating optimization with respect to each one of the involved variables individually and then, it updates the corresponding Lagrange multipliers via a gradient based rule.
Let us first write $(\mathcal{P}_3)$ in the following equivalent form
\begin{align}
\hspace{0pt}(\mathcal{P}_4): \hspace{2pt}  &\min_{\mathbf{Z},\mathbf{F}_{RF},\mathbf{F}_{BB}} -\log_2\det(\mathbf{I}_{L_s}+\tilde{\mathbf{H}}_{ss}\mathbf{Z}\mathbf{Z}^H\tilde{\mathbf{H}}_{ss}^H) + \mathds{1}_{\mathcal{S}}\{\mathbf{Z}\} + \mathds{1}_{\mathcal{F}}\{\mathbf{F}_{RF}\} \nonumber \\
&s.t. \hspace{10pt} \mathbf{Z} = \mathbf{F}_{RF}\mathbf{F}_{BB}
\label{EQ:ADMM_FORM}
\end{align} 
where $\mathbf{Z}$ is an auxiliary $T_s \times N_{st}$ matrix variable, the set $\mathcal{S}$ for a matrix $\mathbf{A}$ is defined as 
\[\mathcal{S} = \left\{\mathbf{A} \in \mathbb{C}^{T_s \times N_{st}}\ | \ \|\mathbf{A}\|_F^2\ \leq P_{max} \ \& \ \|\mathbf{H}_{ps}\mathbf{A}\|_F^2 \leq I_{max}\right\}\] 
and the indicator function $\mathds{1}_{\mathcal{S}}\left\{\mathbf{A}\right\}$ of sets $\mathcal{S}$, $\mathcal{F}$
is defined as 
\begin{equation}
\mathds{1}_{\mathcal{S}(\mathcal{F})}\{\mathbf{A}\} = \begin{cases}
\mathbf{A}, \ \mathbf{A} \in \mathcal{S} (\mathcal{F}) \\
\infty, \ \mathbf{A} \notin \mathcal{S} (\mathcal{F})  
\end{cases}.
\label{EQ:INDI}
\end{equation}
The augmented Lagrangian function of $(\mathcal{P}_4)$ is given by 
\begin{align}
\mathcal{L}_T(\mathbf{Z},\mathbf{F}_{RF},\mathbf{F}_{BB},\mathbf{\Lambda}) =& -\log_2(\mathbf{I}_{L_s}+\tilde{\mathbf{H}}_{ss}\mathbf{Z}\mathbf{Z}^H\tilde{\mathbf{H}}_{ss}^H) + \mathds{1}_{\mathcal{S}}\{\mathbf{Z}\} + 
\mathds{1}_{\mathcal{F}}\{\mathbf{F}_{RF}\} +   \nonumber \\
&\langle\mathbf{\Lambda},\mathbf{Z}-\mathbf{F}_{RF}\mathbf{F}_{BB}\rangle +
\frac{\alpha}{2}\|\mathbf{Z}-\mathbf{F}_{RF}\mathbf{F}_{BB}\|_F^2,
\label{EQ:AUGLAN} 
\end{align}
where $\langle \mathbf{A}, \mathbf{B} \rangle = \sum_{m,n} \mathbf{A}(m,n)\mathbf{B}(m,n)$, for two matrices $\mathbf{A}$ and $\mathbf{B}$, $\mathbf{\Lambda}$ is the $T_s\times N_{st}$ Lagrange Multiplier matrix and $\alpha$ is a scalar penalty parameter.

Following the ADMM approach, the solution to  $(\mathcal{P}_4)$ is given by the following alternating minimization steps
\allowdisplaybreaks
\begin{align}
\label{EQ:Zn}
&\hspace{-1pt}(\mathcal{P}_{4A}): \hspace{1pt} \mathbf{Z}_n = \arg \min_{\mathbf{Z}} \mathcal{L}_T(\mathbf{Z},\mathbf{F}_{RF(n-1)},\mathbf{F}_{BB(n-1)},\mathbf{\Lambda}_{n-1}) \\
\label{EQ:FRFn}
&\hspace{-1pt}(\mathcal{P}_{4B}): \hspace{1pt} \mathbf{F}_{RF(n)} = \arg \min_{\mathbf{F}_{RF}}\mathcal{L}_T(\mathbf{Z}_n,\mathbf{F}_{RF},\mathbf{F}_{BB(n-1)},\mathbf{\Lambda}_{n-1}) \\
\label{EQ:FBBn}
&\hspace{-1pt}(\mathcal{P}_{4C}): \hspace{1pt} \mathbf{F}_{BB(n)} = \arg \min_{\mathbf{F}_{BB}}\mathcal{L}_T(\mathbf{Z}_n,\mathbf{F}_{RF(n)},\mathbf{F}_{BB},\mathbf{\Lambda}_{n-1}) \\
\label{EQ:LAn}
&\hspace{-1pt}(\mathcal{P}_{4D}): \hspace{1pt} \mathbf{\Lambda}_{n} = \arg \min_{\mathbf{\Lambda}}\mathcal{L}_T(\mathbf{Z}_n,\mathbf{F}_{RF(n)},\mathbf{F}_{BB(n)},\mathbf{\Lambda})
\end{align}
where $n$ is the iteration index. 
Let us first derive the solution of $(\mathcal{P}_{4A})$. 
%
Unfortunately, the solution to this problem does not admit a closed form due to its cost function and the constraints on the feasible solution set. However, it is a convex problem (with respect to the optimizing matrix parameter $\mathbf{Z}$). Therefore, a projected gradient based approach will be developed in the following for its solution. The gradient of the cost function with respect to $\mathbf{Z}$ is given by  
\begin{align}
\nabla_\mathbf{Z} \mathcal{L}_T(\mathbf{Z},\mathbf{F}_{RF(n-1)},\mathbf{F}_{BB(n-1)}, \mathbf{\Lambda}_{n-1}) =&
-\frac{2}{\ln(2)}\tilde{\mathbf{H}}_{ss}^H\left(\mathbf{I}_{L_s}+\tilde{\mathbf{H}}_{ss}\mathbf{Z}\mathbf{Z}^H\tilde{\mathbf{H}}_{ss}^H\right)^{-1}\tilde{\mathbf{H}}_{ss}\mathbf{Z} + 
\mathbf{\Lambda}_{n-1} \nonumber \\
&+\alpha\left(\mathbf{Z}-\mathbf{F}_{RF(n-1)}\mathbf{F}_{BB(n-1)}\right)
\label{EQ:GRAD_Z}
\end{align}

Then, the projected gradient descent based solution $\mathbf{Z}_n$ is calculated by updating at each step the solution $\mathbf{Z}_{i,n}$ based on the previous one $\mathbf{Z}_{i-1,n}$ as,
\begin{align}
\mathbf{Z}_{i,n} &= {\Pi}_{\mathcal{S}}\left\{\mathbf{Z}_{i-1,n}- \right. \left.\mu\nabla_\mathbf{Z} \mathcal{L}_T(\mathbf{Z}_{i-1, n},\mathbf{F}_{RF(n-1)},\mathbf{F}_{BB(n-1)}, \mathbf{\Lambda}_{n-1})\right\},
\label{EQ:PG_Z}
\end{align}
where $\mu$ is a step size parameter and ${\Pi}_{\mathcal{S}}$ is the projection onto the set $\mathcal{S}$ operator that can be found by solving the following optimization problem
\begin{align}
\hspace{-25pt}(\mathcal{P}_{5}): \hspace{25pt} &\min_{\mathbf{A}_{\mathcal{S}}}\|\mathbf{A}_{\mathcal{S}}-\mathbf{A}\|_F^2 \nonumber \\
&s.t. \ \mathbf{A}_{\mathcal{S}} \in \mathcal{S},
\label{EQ:PROG_OPTI}
\end{align}
where $\mathbf{A}$ is an arbitrary matrix and $\mathbf{A}_{\mathcal{S}}$ is its projection onto the set $\mathcal{S}$.
The Karush-Kuhn-Tucker (KKT) conditions for $(\mathcal{P}_{5})$ are given by,
\begin{align}
\label{EQ:KKT_PS1}
\mathbf{A}_\mathcal{S}^* - \mathbf{A} + \lambda_1\mathbf{A}_\mathcal{S}^* + \lambda_2\mathbf{H}_{ps}^H\mathbf{H}_{ps}\mathbf{A}_\mathcal{S}^* &= \mathbf{0} \\
\|\mathbf{A}_\mathcal{S}^*\|_F^2 -P_{max} &\leq 0 \\
\|\mathbf{H}_{ps}\mathbf{A}_\mathcal{S}^*\|_F^2 - I_{max} &\leq 0 \\
\lambda_1\left(\|\mathbf{A}_\mathcal{S}^*\|_F^2 -P_{max}\right) &= 0 \\
\lambda_2\left(\|\mathbf{H}_{ps}\mathbf{A}_\mathcal{S}^*\|_F^2 - I_{max}\right) &= 0 \\
\lambda_1, \lambda_2 & \geq 0,
\label{EQ:KKT_PS2}
\end{align}
where $\lambda_1$ and $\lambda_2$ are the corresponding multipliers and $\mathbf{A}_\mathcal{S}^*$ is the optimal solution. From \eqref{EQ:KKT_PS1}-\eqref{EQ:KKT_PS2} it can be shown after some manipulations that operator $\Pi_\mathcal{S}\{\cdot\}$ is given by
\begin{align}
\Pi_{\mathcal{S}}\{\mathbf{A}\} = 
\begin{cases} 
\mathbf{A}, \ &\mathbf{S} \in \mathcal{S} \\
\left[(1+\lambda_1)\mathbf{I}_{L_s}+\lambda_2\mathbf{H}_{ps}^H\mathbf{H}_{ps}\right]^{-1}\mathbf{A}, \ & \mathbf{S} \notin \mathcal{S} 
\end{cases}
\label{EQ:S_PROJ}
\end{align} 
where $\lambda_1$ and $\lambda_2$ are set such that the power ($P_{max}$) and interference constraints ($I_{max}$) are met, i.e. via a bisection method \cite{AGL}.  

The iterations of \eqref{EQ:PG_Z} are running until the following termination criterion is met
\begin{equation}
\|\mathbf{Z}_{i,n}-\mathbf{Z}_{i-1,n}\|_F^2 < \epsilon^{gd},
\label{EQ:TERM_Z}
\end{equation}
where $\epsilon^{gd}$ is a pre-defined tolerance.

Let us move now to the derivation of the solution to $(\mathcal{P}_{4B})$ which can be written in the following simplified form,
\begin{align}
\mathbf{F}_{RF(n)} = \arg \max_{\mathbf{F}_{RF}} \mathds{1}_{\mathcal{F}}\{\mathbf{F}_{RF}\} +  \langle\mathbf{\Lambda}_{n-1},\mathbf{Z}_n-\mathbf{F}_{RF}\mathbf{F}_{BB(n-1)}\rangle + \frac{\alpha}{2}\|\mathbf{Z}_n-\mathbf{F}_{RF}\mathbf{F}_{BB(n-1)}\|_F^2
\end{align}
The solution to this problem can be given by solving the unconstrained problem and then projecting onto the set $\mathcal{F}$. That is,
\begin{align}
\tilde{\mathbf{F}}_{RF(n)} &= \frac{1}{\alpha}\left(\mathbf{\Lambda}_{n-1}+\alpha\mathbf{Z}_n\right)\mathbf{F}_{BB(n)}^H\left(\mathbf{F}_{BB(n)}\mathbf{F}_{BB(n)}^H\right)^{-1}\nonumber  \\ 
\mathbf{F}_{RF(n)} &= \Pi_\mathcal{F}\left\{\tilde{\mathbf{F}}_{RF(n)}\right\},
\label{EQ:FRF_SOL}
\end{align}
where ${\Pi}_{\mathcal{F}}$ is the projection onto the set $\mathcal{F}$ operator that can be found by solving the following optimization problem
\begin{align}
\hspace{-25pt}(\mathcal{P}_{6}): \hspace{25pt} &\min_{\mathbf{A}_{\mathcal{F}}}\|\mathbf{A}_{\mathcal{F}}-\mathbf{A}\|_F^2 \nonumber \\
&s.t. \ \mathbf{A}_{\mathcal{F}} \in \mathcal{F},
\label{EQ:PROJ_OPTII}
\end{align}
where $\mathbf{A}$ is an arbitrary matrix and $\mathbf{A}_{\mathcal{F}}$ is its projection onto the set $\mathcal{F}$. It is straightforward to see that the solution to $(\mathcal{P}_{6})$ is given by the phase of the complex elements of $\mathbf{A}$, that is for $\mathbf{A}_\mathcal{F}  = \Pi_\mathcal{F}\{\mathbf{A}\}$ we have
\begin{align}
\mathbf{A}_\mathcal{F}(m,n) = \begin{cases}
0, \ &\mathbf{A}(m,n) = 0 \\
\frac{\mathbf{A}(m,n)}{\left|\mathbf{A}(m,n)\right|}, \ &\mathbf{A}(m,n) \neq 0 
\end{cases},  
\label{EQ:PROJ_F}
\end{align}
where $\mathbf{A}_\mathcal{F}(m,n)$ and $\mathbf{A}(m,n)$ are the elements at the $m$th row - $n$th column of matrices $\mathbf{A}_\mathcal{F}$ and $\mathbf{A}$ respectively and $|\cdot|$ is the modulus of a complex number.

We move now to the solution of $(\mathcal{P}_{4C})$ which can be written as
\begin{align}
\mathbf{F}_{BB(n)} = \arg \max_{\mathbf{F}_{BB}} \langle\mathbf{\Lambda}_{n-1},\mathbf{Z}_n-\mathbf{F}_{RF(n)}\mathbf{F}_{BB}\rangle + \frac{\alpha}{2}\|\mathbf{Z}_n-\mathbf{F}_{RF(n)}\mathbf{F}_{BB}\|_F^2
\label{EQ:FBB_OPT}
\end{align}
By equating the gradient of \eqref{EQ:FBB_OPT} to zero, we can show that the solution to the corresponding optimization problem admits the following closed form
\begin{equation}
\mathbf{F}_{BB(n)} = \frac{1}{\alpha}\left(\mathbf{F}_{RF(n)}^H\mathbf{F}_{RF(n)}\right)^{-1}\mathbf{F}_{RF(n)}^H(\mathbf{\Lambda}_{n-1} + \alpha \mathbf{Z}_n).
\label{EQ:FBB_SOL}
\end{equation}
Finally, for the Lagrange multiplier ($\mathcal{P}_{4D}$) we use a gradient update rule, that is
\begin{equation}
\mathbf{\Lambda}_{n} = \mathbf{\Lambda}_{n-1} + \alpha\left(\mathbf{Z}_n-\mathbf{F}_{RF(n)}\mathbf{F}_{BB(n)}\right).
\label{EQ:LnGR} 
\end{equation}

According to the termination criteria and the convergence properties of the ADMM sequence discussed later in this sub-section, the primal feasibility condition of $(\mathcal{P}_4)$ is satisfied in the sense that  $\left\|\mathbf{Z}_n - \mathbf{F}_{RF(n)}\mathbf{F}_{BB(n)} \right\|_F \rightarrow 0$. Thus, it is possible for the solution given by the ADMM sequence to violate the power and interference constraints of the original problem $(\mathcal{P}_3)$. To tackle this, upon convergence of the ADMM sequence, we project the digital pre-coder solution to the set $\mathcal{S}'$ defined as,
\begin{align*}
\mathcal{S}' = \left\{\mathbf{A} \in \mathbb{C}^{T_s \times N_{st}}\ | \ \|\mathbf{F}_{RF(\dagger)}\mathbf{A}\|_F^2\ \leq P_{max}\right. \left.\|\mathbf{H}_{ps}\mathbf{F}_{RF(\dagger)}\mathbf{A}\|_F^2 \leq I_{max}\right\}
\end{align*}        
where $\mathbf{F}_{RF(\dagger)}$ is the solution for the analog pre-coder provided by the ADMM sequence and $\mathbf{A}$ is again, an arbitrary matrix. Following a similar procedure to the one that used to derive the projection onto the set $\mathcal{S}$ \eqref{EQ:PROG_OPTI}-\eqref{EQ:S_PROJ}, it can be shown that for $\mathbf{A} \notin \mathcal{S'}$ the projection onto $\mathcal{S}'$ is given by
%
%
%
\begin{align}
{\Pi}_{\mathcal{S}'}\{\mathbf{A}\} = \left[\mathbf{I}_{L_s}+\mathbf{F}_{RF(\dagger)}^H\left(\gamma_1\mathbf{I}_{T_s}+\gamma_2\mathbf{H}_{ps}^H\mathbf{H}_{ps}\right)\mathbf{F}_{RF(\dagger)}\right]^{-1}\mathbf{A},
\label{EQ:PROJJ}
\end{align} 
where the Lagrange Multipliers $\gamma_1$ and $\gamma_2$ are again set such that the power ($P_{max}$) and interference constraints ($I_{max}$) are met.

We may now present some results regarding the convergence of the proposed algorithm under mild conditions. Note that in literature so far, strong convergence results for the ADMM have been derived for convex problems that involve only two blocks of variables. Moreover, strong convergence results for non-convex problems are in general unknown and also an open research problem. Problem $(\mathcal{P}_4)$ involves three blocks of variables and on top of that, it is non-convex. Thus, obtaining strong convergence results is an intractable task and beyond the scopes of the present paper. Here, the following theorem regarding the convergence of the ADMM sequence \eqref{EQ:Zn}-\eqref{EQ:LAn} to an optimal point is given.     
\\
\textit{Theorem 1}: Let $\left\{\mathbf{\Theta}_n\right\} = \left\{\left(\mathbf{Z}_n, \mathbf{F}_{RF(n)}, \mathbf{F}_{BB(n)}, \mathbf{\Lambda}_{n} \right)\right\}$ is a sequence generated by the alternating minimization steps \eqref{EQ:Zn}-\eqref{EQ:LAn}. Let us further assume that the multiplier sequence $\{\mathbf{\Lambda}_{n}\}$ is bounded and satisfies
\begin{equation}
\sum_{n=0}^{\infty}\|\mathbf{\Lambda}_n-\mathbf{\Lambda}_{n-1}\|_F^2 <\infty, 
\label{EQ:LAG_CONV}
\end{equation} 
then, the sequence $\{\mathbf{\Theta}_n\}$ converges always to an optimal point of $(\mathcal{P}_4)$.
\begin{proof}
The proof is given in Appendix A.
\end{proof} 
In the final part of this section, the implementation aspects of the proposed approach are discussed. The core of the developed technique is the four ADMM alternating minimization problems $(\mathcal{P}_{4A})$-$(\mathcal{P}_{4D})$. Matrices $\mathbf{Z}_n$, $\mathbf{F}_{RF(n)}$ and $\mathbf{F}_{BB(n)}$ are initialized with random values. The Lagrange multiplier matrix $\mathbf{\Lambda}_n$ is initialized with zeros. For the termination criteria of this alternating minimization procedure we propose the ones given by
\begin{align}
\left\|\mathbf{Z}_{n} - \mathbf{Z}_{n-1} \right\|_F \leq \epsilon^z \ \& \ 
\left\|\mathbf{Z}_n - \mathbf{F}_{RF(n)}\mathbf{F}_{BB(n)}\right\|_F \leq \epsilon^p
\label{EQ:TERM}
\end{align}  
where $\epsilon^z$ and $\epsilon^p$ are the corresponding tolerances. The first termination criterion guarantees the convergence of variable $\mathbf{Z}_n$ and further the convergence of $(\mathcal{P}_{4A})$ to its optimal value. The second one guarantees that the primal feasibility condition of $(\mathcal{P}_{4})$ is satisfied. As it can be seen by the optimality conditions of $(\mathcal{P}_{4})$, the convergence of the previous two quantities imply the convergence of the ADMM sequence to an optimal point of $(\mathcal{P}_{4})$ (check also the proof of Theorem 1 at the Appendix). 
\begin{algorithm}
\footnotesize
    \caption{Optimal Hybrid Pre-coding Matrix Design}
  \begin{algorithmic}[1]
    \STATE Initialize $\mathbf{Z}_0$, $\mathbf{F}_{RF(0)}$, $\mathbf{F}_{BB(0)}$ with random values and $\mathbf{\Lambda}_{0}$ with zeros   
    \WHILE{The termination criteria of \eqref{EQ:TERM} are not met or $n \leq N_{max}$}
      	\STATE $\mathbf{Z}_{0,n} \gets \mathbf{Z}_{n-1} $
      			\WHILE {$\|\mathbf{Z}_{i,n}-\mathbf{Z}_{i-1,n}\|_F^2\leq \epsilon^{gd}$}  						
      					\STATE Update $\mathbf{Z}_{i,n}$ from \eqref{EQ:PG_Z} 	
      			\ENDWHILE
      	\STATE $\mathbf{Z}_{n} \gets \mathbf{Z}_{i,n}$ 		
      	\STATE Update $\mathbf{F}_{RF(n)}$, $\mathbf{F}_{BB(n)}$ and $\mathbf{\Lambda}_{n}$ from \eqref{EQ:FRF_SOL}-\eqref{EQ:PROJ_F}, \eqref{EQ:FBB_SOL} and \eqref{EQ:LnGR} 
      \ENDWHILE
    \RETURN $\mathbf{F}_{RF(n)}$, $\Pi_{\mathcal{S}'}\left\{\mathbf{F}_{BB(n)}\right\}$
  \end{algorithmic}
\end{algorithm}
Note that due to the non-convex nature of the latter problem, it is also useful to add a termination criterion related to the maximum permitted number of iterations of the ADMM sequence which here is defined as $N_{max}$.       

We close this subsection by discussing the implementation aspects of $(\mathcal{P}_{4A})$ which is solved by the iterative projected gradient technique of \eqref{EQ:PG_Z}. The required complexity can be reduced by smooth starting the gradient descent part of the ADMM's $n$th iteration by setting $\mathbf{Z}_{0,n-1} = \mathbf{Z}_{n-1}$. Furthermore, the tolerance parameter $\epsilon^{gd}$ can be set to a larger value at the beginning of the ADMM sequence and decrease its value 
as the latter converges for better accuracy \cite{BER}\cite{ADMM}. A good rule of thumb is to decrease tolerance $\epsilon^{gd}$ by a power of 10 every time the quantities of \eqref{EQ:TERM} reach its value.  
The complete procedure is given for reference in Algorithm 1.      

\subsection{Hybrid Receiver Design}

Let us now move to the derivation of the post-coding matrix. For complexity issues, here we assume a typical linear receiver based on the MMSE criterion, as it was also discussed in Section III. Based on the constraints related to the structure of the phase only analog counterpart, we may express the corresponding optimization problem as
\begin{align}
\hspace{-25pt}(\mathcal{P}_{7}):\hspace{25pt} &\min_{\mathbf{W}_{RB},\mathbf{W}_{BB}} \mathbb{E}\left\{\|\mathbf{x}_{s}-\mathbf{W}_{BB}^H\mathbf{W}_{RF}^H\mathbf{y}_s\|_F^2\right\} \nonumber  \\
&s.t. \hspace{10pt} \mathbf{W}_{RF} \in \mathcal{F}.
\label{EQ:SU_OPT_AN_W1}
\end{align}
In the absence of hardware limitations that pose the phase only constraints on the receiver, one may find the corresponding optimal digital only post-coder by solving again the unconstrained  problem of \eqref{EQ:MMSE} given that the hybrid pre-coding solutions of Algorithm 1 are applied on the transmitter side. Thus,
by denoting with $\mathbf{W}_D$ the optimal digital only MMSE post-coder, it is straightforward to see from \eqref{EQ:SU_OPT_DIG_W1} that,
\begin{align}
\mathbf{W}_{D}^H =
\left(\mathbf{H}_{ss}\mathbf{F}_{RF}\mathbf{F}_{BB}\mathbf{F}_{BB}^H\mathbf{F}_{RF}^H\mathbf{H}_{ss}^H+\tilde{\mathbf{H}}_{sp}\tilde{\mathbf{H}}_{sp}^H+\sigma_n^2\mathbf{I}_{R_s}\right)^{-1}\mathbf{H}_{ss}\mathbf{F}_{RF}\mathbf{F}_{BB},
\label{EQ:W_D}
\end{align}
where we used the following equations
\begin{align}
\label{EQ:EXP1}
\mathbb{E}\left\{\mathbf{y}_{s}\mathbf{y}_{s}^H\right\} =& \mathbf{H}_{ss}\mathbf{F}_{RF}\mathbf{F}_{BB}\mathbf{F}_{BB}^H\mathbf{F}_{RF}^H\mathbf{H}_{ss}^H + \tilde{\mathbf{H}}_{sp}\tilde{\mathbf{H}}_{sp}^H+\sigma_n^2\mathbf{I}_{R_s} \\
\mathbb{E}\{\mathbf{x}_s\mathbf{y}_s^H\} =& \mathbf{F}_{BB}^H\mathbf{F}_{RF}^H\mathbf{H}_{ss}^H.
\label{EQ:EXP2}
\end{align}

Returning now to the non-convex problem $(\mathcal{P}_{7})$, we apply the methodology of \cite{ARAPH}, \cite{YONINA} related to the design of linear MMSE estimators with complex structural constraints in order to derive the following equivalent form,
\begin{align}
&\min_{\mathbf{W}_{RB},\mathbf{W}_{BB}} \left\|\mathbb{E}\left\{\mathbf{y}_{s}\mathbf{y}_{s}^H\right\}^{1/2}\left(\mathbf{W}_D - \mathbf{W}_{RF}\mathbf{W}_{BB}\right)\right\|_F^2 \nonumber  \\
&s.t. \hspace{10pt} \mathbf{W}_{RF} \in \mathcal{F}.
\label{EQ:W_HYB}
\end{align}
That is, the optimal Hybrid MMSE can be found as the weighted projection of the unconstrained optimal digital MMSE onto the set of post-coders that admit the form $\mathbf{W}_{RF}\mathbf{W}_{BB}$ with $\mathbf{W}_{RF} \in \mathcal{F}$. The proposed solution is based again on the ADMM by expressing the optimization problem of \eqref{EQ:W_HYB} in the following form
\begin{align}
\hspace{-25pt}(\mathcal{P}_8): \hspace{25pt}
&\min_{\mathbf{G},\mathbf{W}_{RF},\mathbf{W}_{BB}} \left\|\mathbb{E}\left\{\mathbf{y}_{s}\mathbf{y}_{s}^H\right\}^{1/2}\left(\mathbf{W}_D - \mathbf{G}\right)\right\|_F^2 + \mathds{1}_{\mathcal{F}}\{\mathbf{W}_{RF}\} \nonumber \\
&s.t. \hspace{10pt} \mathbf{G} = \mathbf{W}_{RF}\mathbf{W}_{BB},
\label{EQ:ADMM_FORM_W}
\end{align} 
where $\mathbf{G}$ is an auxiliary $R_s \times N_{sr}$ matrix variable and the indicator function of set  $\mathcal{F}$ is defined in \eqref{EQ:INDI}.   

The augmented Lagrangian function of $(\mathcal{P}_8)$ is given by 
\begin{align}
\mathcal{L}_R(\mathbf{G},\mathbf{W}_{RF},\mathbf{W}_{BB},\mathbf{\Pi}) =& \left\|\mathbb{E}\left\{\mathbf{y}_{s}\mathbf{y}_{s}^H\right\}^{1/2}\left(\mathbf{W}_D - \mathbf{G}\right)\right\|_F^2 +  
\mathds{1}_{\mathcal{F}}\{\mathbf{W}_{RF}\} +  
\langle\mathbf{\Pi},\mathbf{G}-\mathbf{W}_{RF}\mathbf{W}_{BB}\rangle \nonumber \\
&+\frac{\beta}{2}\|\mathbf{G}-\mathbf{W}_{RF}\mathbf{W}_{BB}\|_F^2,
\label{EQ:AUGLAN_R} 
\end{align}
where $\mathbf{\Pi}$ is the $R_s\times N_{sr}$ Lagrange Multiplier matrix and $\beta$ is a scalar penalty parameter.

Following the ADMM approach, the solution to the optimization problems $(\mathcal{P}_8)$ is given by the following alternating minimization steps
\begin{algorithm}
\footnotesize
    \caption{Hybrid Post-coding Matrix Design}
  \begin{algorithmic}[1]
    \STATE Initialize $\mathbf{G}_0$, $\mathbf{W}_{RF(0)}$, $\mathbf{W}_{BB(0)}$ with random values and $\mathbf{\Pi}_{0}$ with zeros
    \WHILE{The termination criteria of \eqref{EQ:TERM2} are not met or $n \leq N_{max}$}
      	\STATE Update $\mathbf{G}_n$, $\mathbf{W}_{RF(n)}$, $\mathbf{W}_{BB(n)}$ and $\mathbf{\Pi}_{n}$ from \eqref{EQ:GSOL}-\eqref{EQ:WnGR}, respectively      
      \ENDWHILE
    \RETURN $\mathbf{W}_{RF(n)}$, $\mathbf{W}_{BB(n)}$
  \end{algorithmic}
\end{algorithm}
\allowdisplaybreaks
\begin{align}
\label{EQ:Gn}
&\hspace{-1pt}(\mathcal{P}_{8A}): \hspace{1pt} 
\mathbf{G}_n = \arg \min_{\mathbf{G}} \mathcal{L}_R(\mathbf{G},\mathbf{W}_{RF(n-1)},\mathbf{W}_{BB(n-1)},\mathbf{\Pi}_{n-1}) \\
\label{EQ:ERFn}
&\hspace{-1pt}(\mathcal{P}_{8B}): \hspace{1pt}
\mathbf{W}_{RF(n)} = \arg \min_{\mathbf{W}_{RF}}\mathcal{L}_R(\mathbf{G}_n,\mathbf{W}_{RF},\mathbf{W}_{BB(n-1)},\mathbf{\Pi}_{n-1}) \\
\label{EQ:WBBn} 
&\hspace{-1pt}(\mathcal{P}_{8C}): \hspace{1pt}
\mathbf{W}_{BB(n)} = \arg \min_{\mathbf{W}_{BB}}\mathcal{L}_R(\mathbf{G}_n,\mathbf{W}_{RF(n)},\mathbf{W}_{BB},\mathbf{\Pi}_{n-1}) \\
\label{EQ:PAn}
&\hspace{-1pt}(\mathcal{P}_{8D}): \hspace{1pt} \mathbf{\Pi}_{n} = \arg \min_{\mathbf{\Pi}}\mathcal{L}_R(\mathbf{G}_n,\mathbf{W}_{RF(n)},\mathbf{W}_{BB(n)},\mathbf{\Pi})
\end{align}
where $n$ is the iteration index. 
Let us now derive the solution to $(\mathcal{P}_{8A})$-$(\mathcal{P}_{8D})$. Problem $(\mathcal{P}_{8A})$ can be directly solved by equating the gradient of the augmented Lagrangian \eqref{EQ:AUGLAN_R} to zero. Thus, we have
\begin{align}
&\mathbf{G}_n = \left(\mathbb{E}\left\{\mathbf{y}_{s}\mathbf{y}_{s}^H\right\}+\beta\mathbf{I}_{R_s}\right)^{-1} \left(\mathbb{E}\left\{\mathbf{y}_{s}\mathbf{y}_{s}^H\right\}\mathbf{W}_D-\mathbf{\Pi}_{n-1}+\beta\mathbf{W}_{RF(n-1)}\mathbf{W}_{BB(n-1)}\right)
\label{EQ:GSOL}
\end{align}
Problem $(\mathcal{P}_{8B})$ is equivalent to $(\mathcal{P}_{4B})$ through which the analog pre-coder $\mathbf{F}_{RF(n)}$ is updated. Therefore, it is straightforward to show that the solution to $(\mathcal{P}_{8B})$  is given by
\begin{align}
\tilde{\mathbf{W}}_{RF(n)} =& \frac{1}{\beta}\left(\mathbf{\Pi}_{n-1}+\beta\mathbf{G}_n\right)\mathbf{W}_{BB(n)}^H\left(\mathbf{W}_{BB(n)}\mathbf{W}_{BB(n)}^H\right)^{-1} \nonumber \\
\mathbf{W}_{RF(n)} =& \Pi_\mathcal{F}\left\{\tilde{\mathbf{W}}_{RF(n)} \right\},
\label{EQ:WRF_SOL}
\end{align}
where the projection operator ${\Pi}_{\mathcal{F}}$ is defined in \eqref{EQ:PROJ_F}.  

In a similar manner, the solution of $(\mathcal{P}_{8C})$ is equivalent to the one of $(\mathcal{P}_{4C})$ for the update of matrix $\mathbf{F}_{BB(n)}$. Thus, it can be shown that 
\begin{equation}
\mathbf{W}_{BB(n)} = \frac{1}{\beta}\left(\mathbf{W}_{RF(n)}^H\mathbf{W}_{RF(n)}\right)^{-1}\mathbf{W}_{RF(n)}^H(\mathbf{\Pi}_{n-1} + \beta\mathbf{G}_n).
\label{EQ:WBB_SOL}
\end{equation}
Finally, for problem $(\mathcal{P}_{4D})$, a gradient update rule is again used, that is
\begin{equation}
\mathbf{\Pi}_{n} = \mathbf{\Pi}_{n-1} + \beta\left(\mathbf{G}_n-\mathbf{W}_{RF(n)}\mathbf{W}_{BB(n)}\right).
\label{EQ:WnGR} 
\end{equation}  
Termination criteria similar to the ones of problems $(\mathcal{P}_{4A})$-$(\mathcal{P}_{4D})$ may be derived under the same arguments. Therefore, the conditions that must be met are
\begin{align}
\left\|\mathbf{G}_{n} - \mathbf{G}_{n-1} \right\|_F \leq \epsilon^g \ \& \ 
\left\|\mathbf{W}_n - \mathbf{W}_{RF(n)}\mathbf{W}_{BB(n)}\right\|_F \leq \epsilon_2^p,
\label{EQ:TERM2}
\end{align}
where $\epsilon^g$ and $\epsilon_2^p$ are the corresponding pre-defined tolerances. 
Convergence analysis results can also be derived by following the proof given in Appendix for Theorem 1 provided that the  multiplier sequence $\{\mathbf{\Pi}_{n}\}$ is bounded and satisfies $\sum_{n=0}^{\infty}\|\mathbf{\Pi}_n-\mathbf{\Pi}_{n-1}\|_F^2 <\infty$. 
We omit any further reference to avoid repetition. The procedure is summarized in Algorithm 2. 
%
%

\section{Hybrid Analog/Digital Pre-coding Design based on the Frobenious Norm Approximation}

As it is evident from Subsection IV.A, problem $(\mathcal{P}_3)$ requires quite high complexity to be solved due to the involved $\log_2\det(\cdot)$ function, since problem $(\mathcal{P}_{4A})$ of the ADMM sequence via which the solution is derived, does not admit a closed form. Thus, its solution is iteratively calculated by the projected gradient technique in every time step of the ADMM sequence, as described in Subsection IV.A. Therefore, in order to improve the complexity requirements, we need to seek for more efficient solutions to tackle this problem.

To that end, the approach of \cite{ARAPH} is followed where the hybrid pre-coder is designed such that the Frobenious norm of its difference to the optimal digital only solution is minimized. For typical point-to point MIMO systems in mmWave band, this approach is shown to perform satisfactory \cite{ARAPH}\cite{YSZL}, while it requires significantly reduced computational complexity. 

Therefore, the hybrid pre-coder is derived as the solution to the optimization problem
\begin{align}
\hspace{-25pt}(\mathcal{P}_9): \hspace{25pt}  &\min_{\mathbf{F}_{RF},\mathbf{F}_{BB}} \left\|\mathbf{F}_D-\mathbf{F}_{RF}\mathbf{F}_{BB}\right\|^2_F \nonumber \\
&s.t. \hspace{10pt} \|\mathbf{F}_{RF}\mathbf{F}_{BB}\|_F^2 \leq P_{max} \nonumber \\
& \hspace{25pt} \|\mathbf{H}_{ps}\mathbf{F}_{RF}\mathbf{F}_{BB}\|_F^2 \leq I_{max} \nonumber \\
& \hspace{25pt} \mathbf{F}_{RF} \in \mathcal{F}
\label{EQ:SU_OPT_ANAL_FII}
\end{align} 
where the optimal digital only solution $\mathbf{F}_D$ is computed by solving $(\mathcal{P}_2)$ in Section III.             

Following once more the ADMM approach, problem $(\mathcal{P}_9)$ can be cast in the following form  
\begin{align}
\hspace{-10pt}(\mathcal{P}_{10}): \hspace{10pt}  &\min_{\mathbf{T},\mathbf{F}_{RF},\mathbf{F}_{BB}} \left\|\mathbf{F}_D - \mathbf{T}\right\|_F^2 + \mathds{1}_{\mathcal{S}}\{\mathbf{T}\} +\mathds{1}_{\mathcal{F}}\{\mathbf{F}_{RF}\} \nonumber \\
&s.t. \hspace{10pt} \mathbf{T} = \mathbf{F}_{RF}\mathbf{F}_{BB},
\label{EQ:ADMM_FORM_FII}
\end{align} 
where $\mathbf{T}$ is an auxiliary $T_s \times N_{st}$ matrix variable.   

The augmented Lagrangian function of $(\mathcal{P}_{10})$ is given by 
\begin{align}
\mathcal{L}_F(\mathbf{T},\mathbf{F}_{RF},\mathbf{F}_{BB},\mathbf{K}) =& \left\|\mathbf{F}_D - \mathbf{T}\right\|_F^2 +  \mathds{1}_{\mathcal{S}}\{\mathbf{T}\} + 
\mathds{1}_{\mathcal{F}}\{\mathbf{F}_{RF}\} + \nonumber \\
&\langle\mathbf{K},\mathbf{T}-\mathbf{F}_{RF}\mathbf{F}_{BB}\rangle +\frac{\delta}{2}\|\mathbf{T}-\mathbf{F}_{RF}\mathbf{F}_{BB}\|_F^2,
\label{EQ:AUGLAN_TII} 
\end{align}
where $\mathbf{K}$ is the corresponding Lagrange Multiplier and $\delta$ is a scalar penalty parameter. The ADMM sequence for the solution of $(\mathcal{P}_{10})$ is given by  
\allowdisplaybreaks
\begin{align}
\label{EQ:Tn}
&\hspace{-1pt}(\mathcal{P}_{10A}): \hspace{1pt}
\mathbf{T}_n = \arg \min_{\mathbf{T}} \mathcal{L}_F(\mathbf{T},\mathbf{F}_{RF(n-1)},\mathbf{F}_{BB(n-1)},\mathbf{K}_{n-1}) \\
\label{EQ:FRFnII}
&\hspace{-1pt}(\mathcal{P}_{10B}): \hspace{1pt}  
\mathbf{F}_{RF(n)} = \arg \min_{\mathbf{F}_{RF}}\mathcal{L}_F(\mathbf{T}_n,\mathbf{F}_{RF},\mathbf{F}_{BB(n-1)},\mathbf{K}_{n-1}) \\
\label{EQ:FBBnII}
&\hspace{-1pt}(\mathcal{P}_{10C}): \hspace{1pt}
\mathbf{F}_{BB(n)} = \arg \min_{\mathbf{F}_{BB}}\mathcal{L}_F(\mathbf{T}_n,\mathbf{F}_{RF(n)},\mathbf{F}_{BB},\mathbf{K}_{n-1}) \\
\label{EQ:KAn}
&\hspace{-1pt}(\mathcal{P}_{10D}): \hspace{1pt} \mathbf{K}_{n} = \arg \min_{\mathbf{K}}\mathcal{L}_F(\mathbf{T}_n,\mathbf{F}_{RF(n)},\mathbf{F}_{BB(n)},\mathbf{K})
\end{align}
Problem $(\mathcal{P}_{10A})$ can be solved by setting the gradient of the augmented Lagrangian \eqref{EQ:AUGLAN_TII} with respect the variable $\mathbf{T}$ to zero and then projecting the result onto the set $\mathcal{S}$. That is,
\begin{equation}
\mathbf{T}_n = \Pi_\mathcal{S}\left\{\frac{1}{\delta + 1}\left(\mathbf{F}_D - \mathbf{K}_{n-1} + \delta\mathbf{F}_{RF(n-1)}\mathbf{F}_{BB(n-1)}\right)\right\}.
\label{EQ:TSOL}
\end{equation}
Moving on, problems $(\mathcal{P}_{10B})-(\mathcal{P}_{10C})$ are equivalent to $(\mathcal{P}_{8B})-(\mathcal{P}_{8C})$ ones. Thus, in a similar manner it can be shown that
\begin{align}
\tilde{\mathbf{F}}_{RF(n)}  =& \frac{1}{\delta}\left(\mathbf{K}_{n-1}+\delta\mathbf{T}_n\right)\mathbf{F}_{BB(n)}^H\left(\mathbf{F}_{BB(n)}\mathbf{F}_{BB(n)}^H \right)^{-1} \nonumber \\
\mathbf{F}_{RF(n)} =& \Pi_\mathcal{F}\left\{\tilde{\mathbf{F}}_{RF(n)}\right\}, 
\label{EQ:FRF_SOLII}
\end{align}
and 
\begin{equation}
\mathbf{F}_{BB(n)} = \frac{1}{\delta}\left(\mathbf{F}_{RF(n)}^H\mathbf{F}_{RF(n)}\right)^{-1}\mathbf{F}_{RF(n)}^H(\mathbf{K}_{n-1} + \delta\mathbf{T}_n).
\label{EQ:FBB_SOLII}
\end{equation}
Finally, the gradient update rule
\begin{equation}
\mathbf{K}_{n} = \mathbf{K}_{n-1} + \delta\left(\mathbf{T}_n-\mathbf{F}_{RF(n)}\mathbf{F}_{BB(n)}\right),
\label{EQ:KnGR} 
\end{equation}  
is used to update the value of the Lagrange Multiplier.

From \eqref{EQ:TSOL}-\eqref{EQ:FBB_SOLII}, it is clear that this approach involves closed form solutions for problems $(\mathcal{P}_{10A})$-$(\mathcal{P}_{10C})$ and thus requires significantly reduced complexity compared to the one of $(\mathcal{P}_4)$. For the receiver side, Algorithm 2 can be again used to provide the hardware constrained MMSE based solution.    

Termination criteria similar to the ones of problems $(\mathcal{P}_{4A})$-$(\mathcal{P}_{4D})$ and  $(\mathcal{P}_{8A})$-$(\mathcal{P}_{8D})$ may be derived under the same arguments. Therefore, the conditions that must be met are
\begin{align}
\left\|\mathbf{T}_{n} - \mathbf{T}_{n-1} \right\|_F \leq \epsilon^{t} \ \& \ 
\left\|\mathbf{T}_n - \mathbf{F}_{RF(n)}\mathbf{F}_{BB(n)}\right\|_F \leq \epsilon_3^{p},
\label{EQ:TERM3}
\end{align}
\begin{algorithm}
\footnotesize
    \caption{Hybrid Pre-coding Matrix Design via Frobenious Norm Approximation}
  \begin{algorithmic}[1]
    \STATE Compute $\mathbf{F}_D$ by solving $(\mathcal{P}_2)$ as described in Sec. III
    \STATE Initialize $\mathbf{T}_0$, $\mathbf{F}_{RF(0)}$, $\mathbf{F}_{BB(0)}$ with random values and $\mathbf{K}_{0}$ with zeros
    \WHILE{The termination criteria of \eqref{EQ:TERM3} are not met or $n \leq N_{max}$}
      	\STATE Update $\mathbf{T}_n$, $\mathbf{F}_{RF(n)}$, $\mathbf{F}_{BB(n)}$ and $\mathbf{K}_{n}$ from \eqref{EQ:TSOL}-\eqref{EQ:KnGR}   		
      \ENDWHILE
    \RETURN $\mathbf{F}_{RF(n)}$, $\Pi_{\mathcal{S}'}\left\{\mathbf{F}_{BB(n)}\right\}$
  \end{algorithmic}
\end{algorithm}
where $\epsilon^{t}$ and $\epsilon_3^{p}$ are the corresponding pre-defined tolerances. 
Convergence analysis results can be derived by assuming that the Lagrange Multiplier sequence $\{\mathbf{K}_n\}$ is bounded and satisfies $\sum_{n=0}^{\infty}\|\mathbf{K}_n-\mathbf{K}_{n-1}\|_F^2 <\infty$, similar to problems $(\mathcal{P}_4)$ and $(\mathcal{P}_8)$. Any further details are omitted in order to avoid repetitions. The procedure is summarized in Algorithm 3. 
%

\section{Simulations}
In this section, numerical results are presented for evaluating the performance of the proposed hybrid approaches. An environment of $N_p = 15$ propagation paths is assumed for all the involved links (SU-to SU, the SU-to-PU and PU-to-SU) and the element spacing on the ULA of each node is set to $d = \frac{\lambda}{2}$. The performance of the proposed approaches is examined in terms of the achieved mean spectral efficiency versus the Signal-to-Noise-Ratio (SNR) over 100 channel realizations. Let us now refer to the parameter tuning of the proposed Algorithms. 
The following numbers are the same for all the experiments presented in this paper. For Algorithm 1, the parameters are set as $\mu = 10^{-3}$, $\alpha = 10$, $\epsilon^z = 10^{-3}$, $\epsilon^p = 10^{-4}$ and $\epsilon^{gd} = 10^{-2}$ (initial value which is updated based on the convergence of the ADMM sequence as described in Section IV.A). 

For Algorithm 2, the parameters are set as  $\beta = 1$, $\epsilon_g = 10^{-3}$ and $\epsilon_2^p = 10^{-4}$. Finally, for Algorithm 3, the parameters are set as $\delta = 10$, $\epsilon_t = 10^{-3}$ and $\epsilon_3^p = 10^{-4}$. The maximum number of iterations per ADMM sequence is set to $N_{max} = 500$ for all the previous Algorithms.
For comparison purposes, we also plot in the following figures a) the performance of the optimal digital only transceiver which is derived as described in Section III, 
and  b) the codebook based approach of \cite{ARAPH} which is adapted to the cognitive scenario proposed here. This technique aims at minimizing the distance of the hybrid beamformer to the optimal digital one by solving a sparse reconstruction problem. 
\begin{figure}[h]
    \centering
    \includegraphics[width=0.7\textwidth]{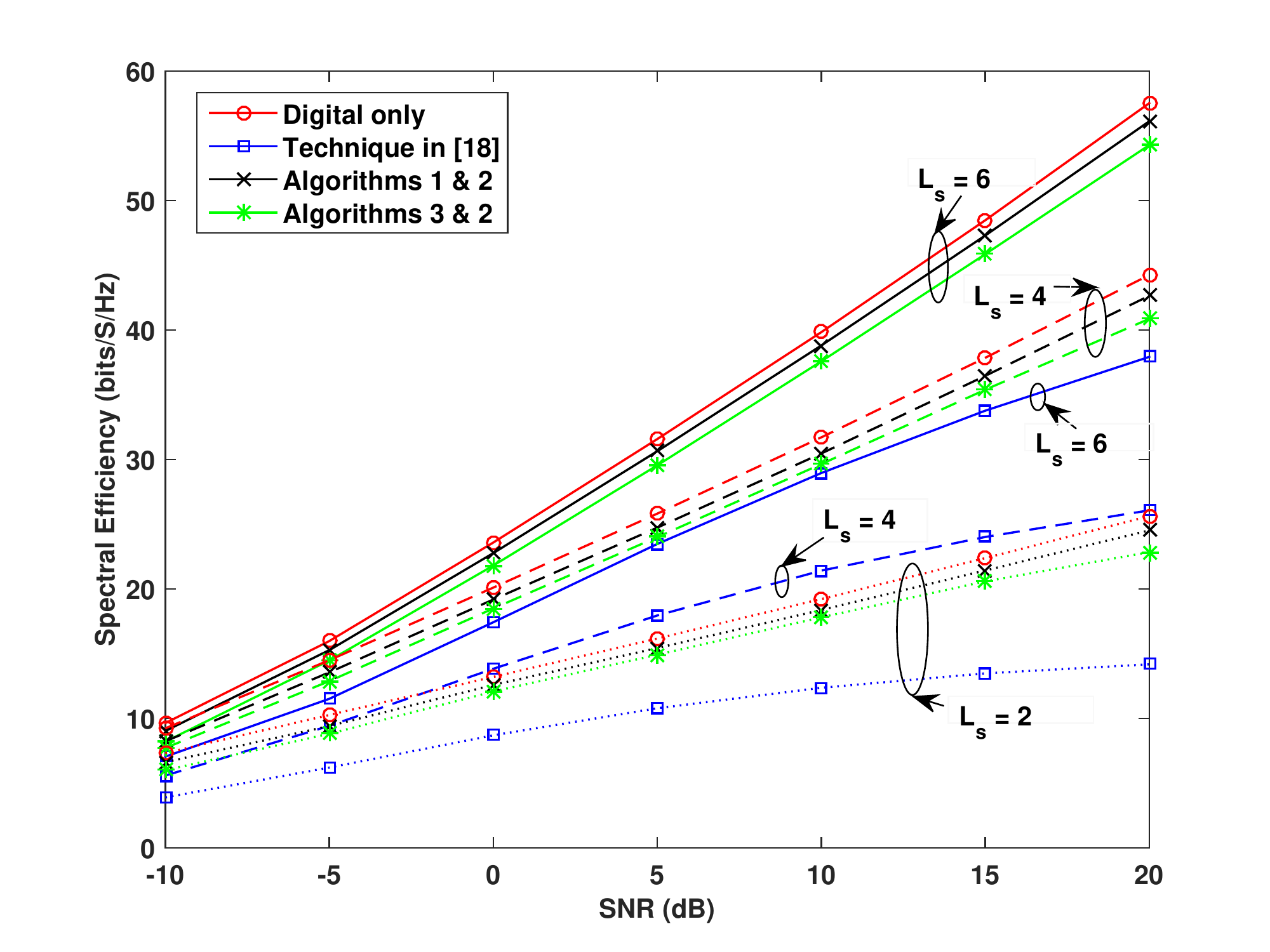}
    \caption{Spectral Efficiency of the Hybrid Pre-coding Techniques vs SNR for a $64 \times 16$ MIMO system and $N_{st} = N_{sr} = L_s = \{2,4,6\}$  }
    \label{FIG:RI}
\end{figure}
The analog counterpart is determined by a codebook and the digital one is computed by applying the well-known Orthogonal Matching Pursuit (OMP) technique. In the cognitive radio setup examined here, the pre-coding part is derived by first applying this approach to approximate the solution of $(\mathcal{P}_2)$ and then,  the digital counterpart is projected onto $\mathcal{S}'$ \eqref{EQ:PROJJ}, so as to force the hybrid solution to satisfy the interference and power constraints. The post-coding matrix at the receiver is derived by applying directly the approach of \cite{ARAPH} to minimize the distance to the optimal MMSE receiver of \eqref{EQ:W_D} by using \eqref{EQ:EXP1}-\eqref{EQ:EXP2} in place of the corresponding values in the original approach.

In Figure 3, the performance of the different hybrid techniques is examined for a SU $64 \times 16$ MIMO system with $N_{st}=N_{sr} = \{2, 4, 6\} $ RF chains. As it was discussed above, the performance of the optimal digital only solution is plotted for $L_s = \{2, 4, 6\}$ for a fair comparison, since the hybrid approaches can support transmission of maximum $\min\{N_{st},N_{sr}\}$ streams. The maximum transmission power and interference constraints are set to $P_{max} = 1$ and $I_{max} = 1$, respectively. The PU is assumed to transmit a signal of rank $L_p = 4$. 

From the results depicted in Figure 3, it is evident that both of the proposed hybrid approaches achieve close performance to the one of the digital only solution for all the examined cases. The one based on the mutual information maximization problem (``Algorithms 1 \& 2'') achieves better performance compared to the one that minimizes the distance of the hybrid pre-coder to digital only solution (``Algorithms 3 \& 2''). Moreover note that as the SNR increases, the gap in the performance of these two approaches increases too. This can be explained by the fact that at low SNR values, the performance is mainly dominated by noise and thus the approximation of the original problem by the solution of Algorithms 3 \& 2 has lesser impact on the system's performance. 
\begin{figure}[h]
    \centering
    \includegraphics[width=0.7\textwidth]{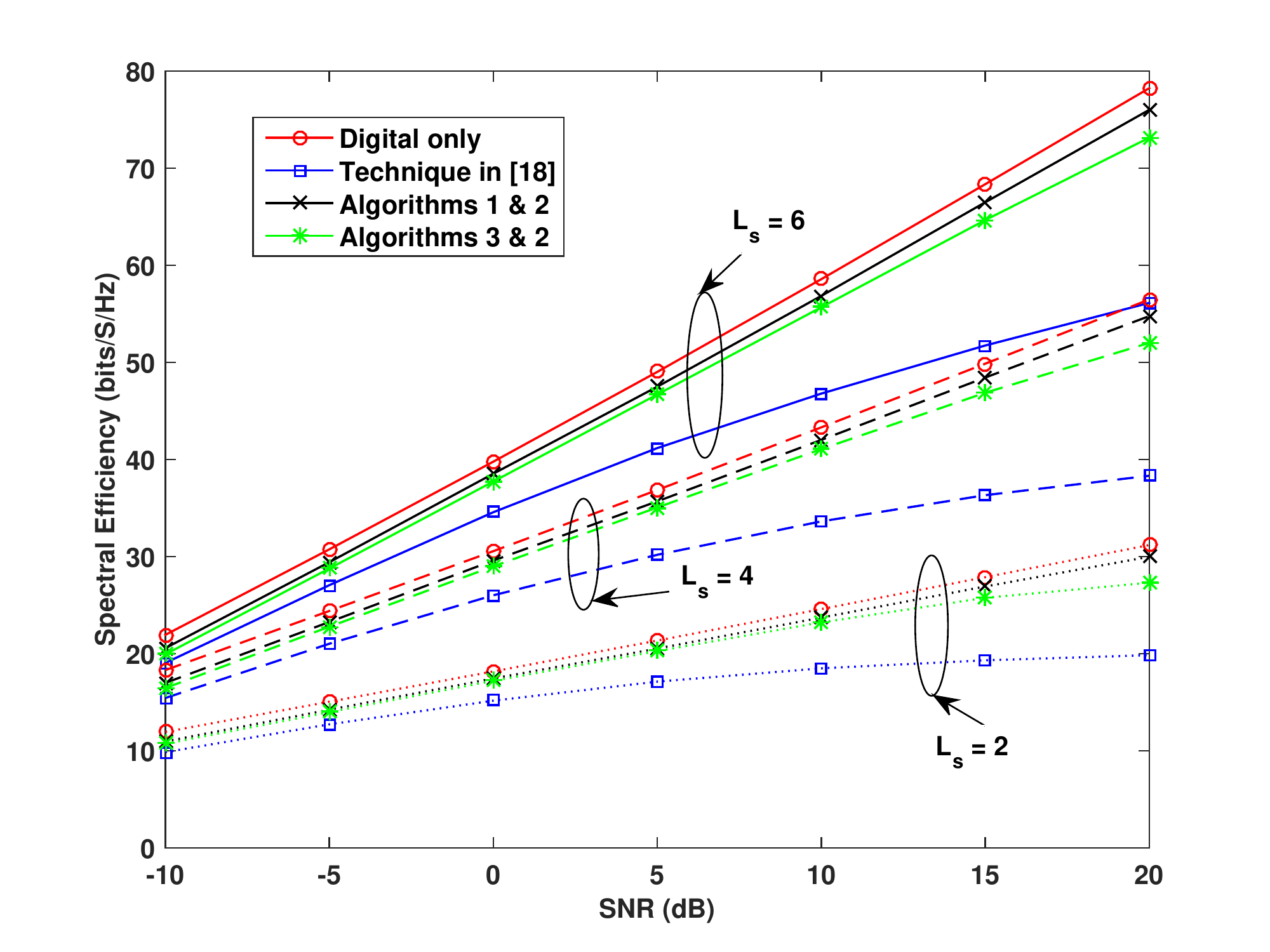}
    \caption{Spectral Efficiency of the Hybrid Pre-coding Techniques vs SNR for a $100 \times 64$ MIMO system and $N_{st} = N_{sr} = L_s = \{2,4,6\}$  }
    \label{FIG:RII}
\end{figure}
On the contrary, for high SNR values, the noise has negligible impact on the performance and thus the degradation due to the aforementioned approximation is more profound. A final point in Figure 3 is related to the performance of the technique in \cite{ARAPH} which is plotted for all the considered cases. As it is evident, the performance of that technique is severely inferior to the proposed ones, especially for high SNR values. This is due to the codebook use that restricts the solution set of the analog counterpart, as it was also observed in \cite{SY} and \cite{YSZL}.

Similar results are also observed in Figure 4 where the experimental set-up is the same with Figure 3, though now a $100\times 64$ SU system is assumed. All the techniques benefit from the larger number of antennas and achieve improved performance. The proposed hybrid techniques achieve again performance very close to the one of the digital only solution for systems with larger number of antenna elements. It is also noteworthy that the performance gap          between the proposed techniques and the one in \cite{ARAPH} is smaller for larger array systems though still quite large, especially for high SNR values.  

In Figure 5, we examine the impact of the PU signal's rank on the performance of the proposed techniques. The number of RF chains is fixed to $N_{st}= N_{sr}= 6$  and  $I_{max} = 1$. We consider PU signals of rank $L_p = \{2, 6, 9\}$ respectively. The rest of the parameters are the same with the ones of the experiments of Figures 3 and 4. As it was expected, the performance of all the techniques is degraded as the rank of PU signal's increases. 
\begin{figure}[h]
    \centering
    \includegraphics[width=0.7\textwidth]{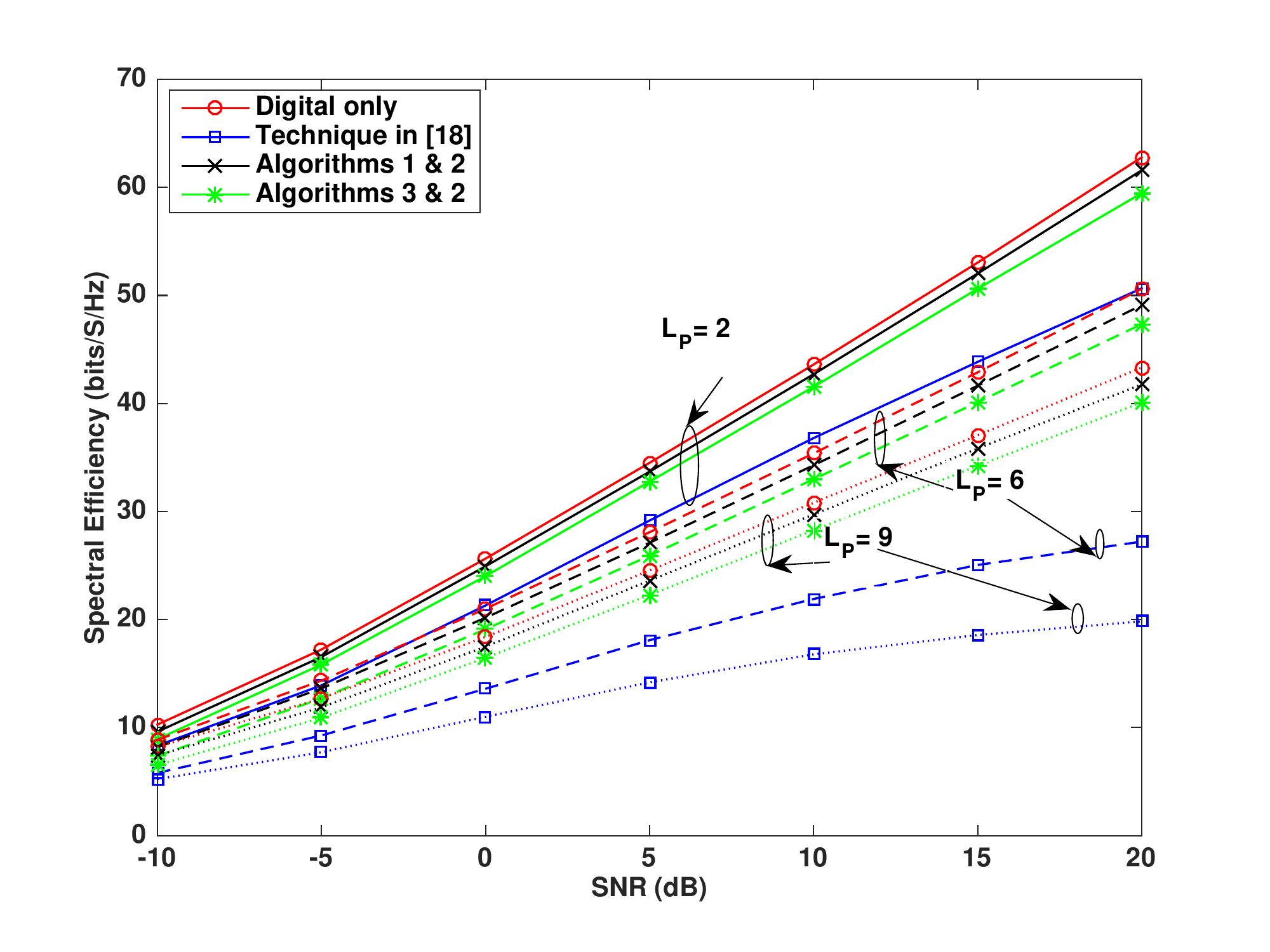}
    \caption{Spectral Efficiency of the Hybrid Pre-coding Techniques vs SNR for a $64 \times 16$ MIMO system and $L_p = \{2,6,9\}$ }
    \label{FIG:RIII}
\end{figure}
This can be attributed to the fact that more of the available degrees of freedom are occupied by the PU, as its signal's rank increases and thus, the SU has to reduce further its transmitted power in order to satisfy the interference temperature constraint. 
Both of the proposed hybrid approaches achieve once again close performance to the one of the digital only optimal solution, independently of the PU signal's rank. On the contrary, the performance of the 
approach in \cite{ARAPH} is degraded with an increase on the PU signal's of rank.

Finally, in Figure 6, the impact of the interference constraint on the performance of the proposed techniques is examined. The experimental set-up is the same with the one of Figure 5, though now the values $I_{max} = \{10, 0.0001\}$ are considered and the PU signal's rank is fixed to $L_p=4$. Under both cases, the proposed hybrid techniques achieve once more close performance to the one of the digital only case. On the other hand, the approach in \cite{ARAPH} exhibits a severe degradation in the performance for low $I_{max}$ values, as it shown in the same figure.

\section{Conclusions}
In this paper, two novel hybrid analog/digital transceiver designs were proposed for CR large-scale antenna systems. The proposed approaches aim at maximizing the spectral efficiency of the SU while
\begin{figure}[h]
    \centering
    \includegraphics[width=0.7\textwidth]{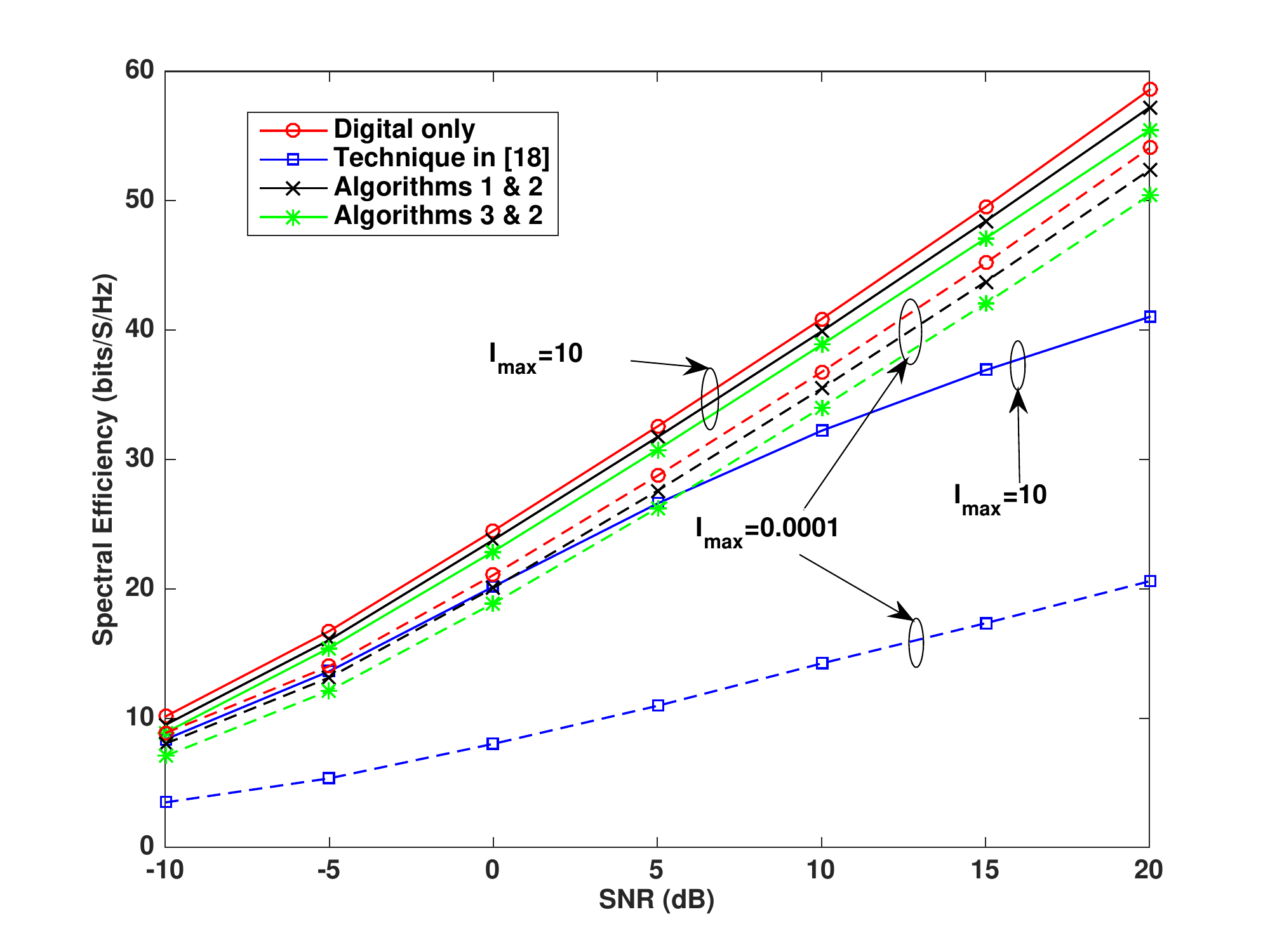}
    \caption{Spectral Efficiency of the Hybrid Pre-coding Techniques vs SNR for a $64 \times 16$ MIMO system and $I_{max} = \{10,0.0001\}$ }
    \label{FIG:RIV}
\end{figure}  
keeping the interference to the PU bellow a pre-defined threshold. Both of the approaches decouple the
transmitter-receiver optimization problem and are based on the so-called ADMM. The first technique derives the optimal pre-coder by maximizing the mutual information between the transceiver's ends. The post-coding matrix at the receiver side is derived by a novel hardware constrained MMSE approach. The second technique aims at reducing the required computational complexity by deriving the optimal pre-coding matrix at the transmitter by minimizing the distance (with respect to the Frobenious norm) of the hybrid solution to the optimal digital one. At the receiver side, the post-coding matrix is based on the same solution to the one of the first technique. The second approach requires reduced complexity as it replaces the highly complex mutual information maximization problem with a simpler least squares one. Simulations show that both the proposed approaches achieve close performance to the one of the corresponding digital only transceiver while having less requirements in hardware and power consumption than the latter. Future works that are currently under development include the extension of the proposed approach to partially connected hybrid solutions and frequency selective channels.           

\appendix[Proof of Theorem 1]
The proof is divided into two parts. At first, we show that given the convergence of the Lagrange Multiplier sequence, as it is defined in \eqref{EQ:LAG_CONV}, the sequence $\{\mathbf{\Theta}_n\}$ has always a limit point. To that end, we begin by showing that the augmented Lagrangian function given in \eqref{EQ:AUGLAN} is bounded from below. The equivalent scaled form \cite{ADMM} of the latter function is given by
\begin{align}
\mathcal{L}_T(\mathbf{Z},\mathbf{F}_{RF},\mathbf{F}_{BB},\mathbf{\Lambda}) &= -\log_2\det(\mathbf{I}_{L_s}+\tilde{\mathbf{H}}_{ss}\mathbf{Z}\mathbf{Z}^H\tilde{\mathbf{H}}_{ss}^H) + \mathds{1}_{\mathcal{S}}\{\mathbf{Z}\} + \mathds{1}_{\mathcal{F}}\{\mathbf{F}_{RF}\} +   \nonumber \\
& +\frac{\alpha}{2}\left\|\mathbf{Z}+\frac{\mathbf{\Lambda}}{\alpha}-\mathbf{F}_{RF}\mathbf{F}_{BB}\right\|_F^2 - \frac{1}{2\alpha}\|\mathbf{\Lambda}\|_F^2.
\label{EQ:AUGLANSCALE} 
\end{align}
The claim then is justified by considering the boundness of $\{\mathbf{\Lambda}_n\}$ in \eqref{EQ:AUGLANSCALE}. Moreover, the optimization problems $(\mathcal{P}_{4A})$, $(\mathcal{P}_{4C})$ and $(\mathcal{P}_{4D})$ are convex with respect the optimizing variables and hence, $\mathbf{Z}_n$, $\mathbf{F}_{BB(n)}$ and $\mathbf{\Lambda}_n$ are the corresponding unique minimizers. Problem $(\mathcal{P}_{4B})$ is non-convex, though set $\mathcal{F}$ is closed and bounded and thus the sequence $\mathbf{F}_{RF(n)}$ converges always to a limit point. Let us further assume that variable $\alpha$ is set to a large value such that the Augmented Lagrangian Function is strongly convex with respect the optimizing variable in all the $(\mathcal{P}_{4A}) - (\mathcal{P}_{4C})$. Due to the aforementioned strong convexity of the Lagrange function the following equations hold \cite{ADMM-NMF}
\begin{align}
\label{EQ:STR_DUAL_START}
&\mathcal{L}_T\left(\mathbf{Z}_{n}, \mathbf{F}_{RF(n)},\mathbf{F}_{BB(n)}, \mathbf{\Lambda}_n \right) - \mathcal{L}_T\left(\mathbf{Z}_{n+1}, \mathbf{F}_{RF(n)},\mathbf{F}_{BB(n)}, \mathbf{\Lambda}_n \right) \geq \nonumber \\
&\left\langle \partial_Z \mathcal{L}_T\left(\mathbf{Z}_{n+1}, \mathbf{F}_{RF(n)},\mathbf{F}_{BB(n)}, \mathbf{\Lambda}_n \right), 
\mathbf{Z}_n-\mathbf{Z}_{n+1}\right\rangle  + \alpha \|\mathbf{Z}_{n} - \mathbf{Z}_{n+1}\|_F^2  \\
&\mathcal{L}_T\left(\mathbf{Z}_{n+1}, \mathbf{F}_{RF(n)},\mathbf{F}_{BB(n)}, \mathbf{\Lambda}_n \right) - \mathcal{L}_T\left(\mathbf{Z}_{n+1}, \mathbf{F}_{RF(n+1)},\mathbf{F}_{BB(n)}, \mathbf{\Lambda}_n \right) \geq \nonumber \\
&\left\langle\partial_{F_{RF}} \mathcal{L}_T\left(\mathbf{Z}_{n+1}, \mathbf{F}_{RF(n+1)},\mathbf{F}_{BB(n)}, \mathbf{\Lambda}_n \right), 
\mathbf{F}_{RF(n)}-\mathbf{F}_{RF(n+1)}\right\rangle  + \alpha \|\mathbf{F}_{RF(n)}-\mathbf{F}_{RF(n+1)}\|_F^2 \\
&\mathcal{L}_T\left(\mathbf{Z}_{n+1}, \mathbf{F}_{RF(n+1)},\mathbf{F}_{BB(n)}, \mathbf{\Lambda}_n \right) - \mathcal{L}_T\left(\mathbf{Z}_{n+1}, \mathbf{F}_{RF(n+1)},\mathbf{F}_{BB(n+1)}, \mathbf{\Lambda}_n \right) \geq \nonumber \\
&\left\langle\partial_{F_{BB}} \mathcal{L}_T\left(\mathbf{Z}_{n+1}, \mathbf{F}_{RF(n+1)},\mathbf{F}_{BB(n+1)}, \mathbf{\Lambda}_n \right), 
\mathbf{F}_{BB(n)}-\mathbf{F}_{BB(n+1)}\right\rangle  + \alpha \|\mathbf{F}_{BB(n)}-\mathbf{F}_{BB(n+1)}\|_F^2,
\label{EQ:STR_DUAL_END}
\end{align}
where $\partial_\mathbf{A}$ denotes the sub-gradient with respect $\mathbf{A}$.
\\
Moreover, since $\mathbf{Z}_{n+1}$, $\mathbf{F}_{RF(n+1)}$ and $\mathbf{F}_{BB(n+1)}$ are the minimizers of $\mathcal{L}_T\left(\mathbf{Z}, \mathbf{F}_{RF(n)},\mathbf{F}_{BB(n)}, \mathbf{\Lambda}_n \right)$, $\mathcal{L}_T\left(\mathbf{Z}_{n+1}, \mathbf{F}_{RF}, \mathbf{F}_{BB(n)}, \mathbf{\Lambda}_n \right)$ and $\mathcal{L}_T\left(\mathbf{Z}_{n+1}, \mathbf{F}_{RF(n+1)},\mathbf{F}_{BB},\mathbf{\Lambda}_n \right)$ respectively, we have
\begin{align}
\label{EQ:OPT_START}
\left\langle\partial_Z \mathcal{L}_T\left(\mathbf{Z}_{n+1}, \mathbf{F}_{RF(n)},\mathbf{F}_{BB(n)}, \mathbf{\Lambda}_n \right), \mathbf{Z}_n-\mathbf{Z}_{n+1} \right\rangle &\geq 0 \\
\left\langle\partial_{F_{RF}} \mathcal{L}_T\left(\mathbf{Z}_{n+1}, \mathbf{F}_{RF(n+1)},\mathbf{F}_{BB(k)}, \mathbf{\Lambda}_n \right), 
\mathbf{F}_{RF(n)}-\mathbf{F}_{RF(n+1)}\right\rangle &\geq 0 \\
\left\langle\partial_{F_{BB}} \mathcal{L}_T\left(\mathbf{Z}_{n+1}, \mathbf{F}_{RF(n+1)},\mathbf{F}_{BB(n+1)}, \mathbf{\Lambda}_n \right), 
\mathbf{F}_{BB(n)}-\mathbf{F}_{BB(n+1)}\right\rangle &\geq 0,
\label{EQ:OPT_END}
\end{align}
By combining \eqref{EQ:STR_DUAL_START}-\eqref{EQ:STR_DUAL_END} with \eqref{EQ:OPT_START}-\eqref{EQ:OPT_END} we may show that   
\begin{align}
&\mathcal{L}_T\left(\mathbf{Z}_{n}, \mathbf{F}_{RF(n)},\mathbf{F}_{BB(n)}, \mathbf{\Lambda}_n \right) -\mathcal{L}_T\left(\mathbf{Z}_{n+1}, \mathbf{F}_{RF(n+1)},\mathbf{F}_{BB(n+1)}, \mathbf{\Lambda}_{n+1}\right) = \\
&\mathcal{L}_T\left(\mathbf{Z}_{n}, \mathbf{F}_{RF(n)},\mathbf{F}_{BB(n)}, \mathbf{\Lambda}_n \right) -\mathcal{L}_T\left(\mathbf{Z}_{n+1}, \mathbf{F}_{RF(n)},\mathbf{F}_{BB(n)}, \mathbf{\Lambda}_{n}\right) +  \nonumber \\
&\mathcal{L}_T\left(\mathbf{Z}_{n+1}, \mathbf{F}_{RF(n)},\mathbf{F}_{BB(n)}, \mathbf{\Lambda}_n \right) -\mathcal{L}_T\left(\mathbf{Z}_{n+1}, \mathbf{F}_{RF(n+1)},\mathbf{F}_{BB(n+1)}, \mathbf{\Lambda}_{n}\right) \geq \\ 
&\alpha\left\|\tilde{\mathbf{\Theta}}_n-\tilde{\mathbf{\Theta}}_{n+1}\right\|_F^2-\frac{1}{\alpha}\left\|{\mathbf{\Lambda}}_n-{\mathbf{\Lambda}}_{n+1}\right\|_F^2,
\end{align}
where $\tilde{\mathbf{\Theta}}_n = \left(\mathbf{Z}_{n}, \mathbf{F}_{RF(n)},\mathbf{F}_{BB(n)}\right)$. 
Taking the summation of the last equality and considering that $\mathcal{L}_T\left(\mathbf{Z}, \mathbf{F}_{RF},\mathbf{F}_{BB}\right)$ is bounded from below we have, 
\begin{equation}
\alpha\sum_{n=0}^\infty\left\|\tilde{\mathbf{\Theta}}_n-\tilde{\mathbf{\Theta}}_{n+1}\right\|_F^2-\frac{1}{\alpha}\sum_{n=0}^\infty\left\|{\mathbf{\Lambda}}_n-{\mathbf{\Lambda}}_{n+1}\right\|_F^2 \leq \infty. 
\end{equation}
Since the second term is bounded by assumption, we can deduce that
\begin{equation}
\sum_{n=0} \|\mathbf{\Theta}_n-\mathbf{\Theta}_{n-1}\|_F < \infty,
\end{equation}
which implies the existence of the limit point of the ADMM sequence in the sense that
\begin{equation}
\mathbf{\Theta}_n-\mathbf{\Theta}_{n-1} \rightarrow \mathbf{0}.
\end{equation}  
We now move to the second part of the proof where the aim is to show that any limit point of the ADMM sequence \eqref{EQ:Zn} - \eqref{EQ:LAn} is a local optimal point of $(\mathcal{P}_4)$. By the KKT conditions, an optimal point $\mathbf{\Theta}_* = \left\{\mathbf{Z}_*, \mathbf{F}_{RF(*)}, \mathbf{F}_{BB(*)}, \mathbf{\Lambda}_{*}\right\}$ of $(\mathcal{P}_4)$ should satisfy the set of equations given by
\begin{align}
\label{P4_OPT_START}
-\tilde{\mathbf{H}}_{ss}^H\left(\mathbf{I}_{L_s} + \tilde{\mathbf{H}}_{ss}\mathbf{Z}_*\mathbf{Z}_*^H\tilde{\mathbf{H}}_{ss}\right)\tilde{\mathbf{H}}_{ss}\mathbf{Z}_* + \mathbf{\Lambda}_* &= \mathbf{0} \\
\mathbf{\Lambda}_*\mathbf{F}_{BB(*)}^H &\in \partial \mathds{1}_\mathcal{F}\left\{\mathbf{F}_{RF(*)}\right\} \\
\mathbf{F}_{RF(*)}^H\mathbf{\Lambda}_* &\in \partial \mathds{1}_\mathcal{S}\left\{\mathbf{F}_{BB(*)}\right\} \\
\mathbf{Z}_* - \mathbf{F}_{RF(*)}\mathbf{F}_{BB(*)} &= \mathbf{0}.
\label{P4_OPT_END}
\end{align}

Now, from the optimality conditions of \eqref{EQ:Zn} - \eqref{EQ:FBBn} we equivalently have
\begin{align}
\label{P4_AUG_LAG_OPT_START}
&-\tilde{\mathbf{H}}_{ss}^H\left(\mathbf{I}_{L_s} + \tilde{\mathbf{H}}_{ss}\mathbf{Z}_n\mathbf{Z}_n^H\tilde{\mathbf{H}}_{ss}\right)\tilde{\mathbf{H}}_{ss}\mathbf{Z}_n + \mathbf{\Lambda}_{n-1} + \alpha\left(\mathbf{Z}_{n} - \mathbf{F}_{RF(n-1)}\mathbf{F}_{BB(n-1)}\right) = \mathbf{0} \\
&-\mathbf{\Lambda}_{n-1}\mathbf{F}_{BB(n-1)}^H -\alpha\left(\mathbf{Z}_{n} - \mathbf{F}_{RF(n-1)}\mathbf{F}_{BB(n-1)}\right)\mathbf{F}_{BB(n-1)}^H \in \partial \mathds{1}_\mathcal{F}\left\{\mathbf{F}_{RF(n)}\right\} \\
&-\mathbf{F}_{RF(n)}^H\mathbf{\Lambda}_{n-1} -\alpha\mathbf{F}_{RF(n)}^H\left(\mathbf{Z}_{n} - \mathbf{F}_{RF(n)}\mathbf{F}_{BB(n)}\right) \in \partial \mathds{1}_\mathcal{S}\left\{\mathbf{F}_{BB(n)}\right\}.
\label{P4_AUG_LAG_OPT_END}
\end{align}

Since convergence of the ADMM sequence is guaranteed, we have that $\left\{\mathbf{\Theta}_n\right\} \rightarrow \mathbf{\Theta}_\dagger$, as $n \rightarrow \infty$, where $\mathbf{\Theta}_\dagger = \left\{\mathbf{Z}_\dagger, \mathbf{F}_{RF(\dagger)}, \mathbf{F}_{BB(\dagger)}, \mathbf{\Lambda}_{\dagger}\right\}$ is the corresponding limit point. Furthermore, as $n \rightarrow \infty$, $\mathbf{\Lambda}_n - \mathbf{\Lambda}_{n-1} \rightarrow \mathbf{0}$ and $\mathbf{Z}_n - \mathbf{F}_{RF(n)}\mathbf{F}_{BB(n)} \rightarrow \mathbf{0} $ from \eqref{EQ:LAG_CONV} and \eqref{EQ:LnGR}, respectively. The previous with their turn imply that
\begin{equation}
\mathbf{Z}_\dagger - \mathbf{F}_{RF(\dagger)}\mathbf{F}_{BB(\dagger)} \rightarrow \mathbf{0}.
\label{EQ:CONV}
\end{equation}
By taking the limit of \eqref{P4_AUG_LAG_OPT_START}-\eqref{P4_AUG_LAG_OPT_END} and applying \eqref{EQ:CONV} in order to simplify the results we get
\begin{align}
\label{P4_AUG_LAG_DAG_START}
&-\tilde{\mathbf{H}}_{ss}^H\left(\mathbf{I}_{L_s} + \tilde{\mathbf{H}}_{ss}\mathbf{Z}_\dagger\mathbf{Z}_\dagger^H\tilde{\mathbf{H}}_{ss}\right)\tilde{\mathbf{H}}_{ss}\mathbf{Z}_\dagger + \mathbf{\Lambda}_{\dagger} + \alpha\left(\mathbf{Z}_{\dagger} - \mathbf{F}_{RF(\dagger)}\mathbf{F}_{BB(\dagger)}\right) = \mathbf{0} \\
&\mathbf{\Lambda}_{\dagger}\mathbf{F}_{BB(\dagger)}^H -\alpha\left(\mathbf{Z}_{\dagger} - \mathbf{F}_{RF(\dagger)}\mathbf{F}_{BB(\dagger)}\right)\mathbf{F}_{BB(\dagger)}^H \in \partial \mathds{1}_\mathcal{F}\left\{\mathbf{F}_{RF(\dagger)}\right\} \\
&\mathbf{F}_{RF(\dagger)}^H\mathbf{\Lambda}_{\dagger} -\alpha\mathbf{F}_{RF(\dagger)}^H\left(\mathbf{Z}_{\dagger} - \mathbf{F}_{RF(\dagger)}\mathbf{F}_{BB(\dagger)}\right) \in \partial \mathds{1}_\mathcal{S}\left\{\mathbf{F}_{BB(\dagger)}\right\}.
\label{P4_AUG_LAG_DAG_END}
\end{align}
By comparing \eqref{EQ:CONV}-\eqref{P4_AUG_LAG_DAG_END} to \eqref{P4_OPT_START}-\eqref{P4_OPT_END}, it is straightforward to see that the limit point $\mathbf{\Theta}_\dagger$ satisfies the KKT conditions and thus it is optimal. \hfill  $\blacksquare$

\bibliographystyle{IEEEtran}

\end{document}